\documentclass[useAMS,usenatbib]{mn2e}
\usepackage{graphicx}
\usepackage{amssymb}
\usepackage{hyperref}

\newcommand{\av}{\alpha_{{\rm v}}}
\newcommand{\bv}{\beta_{{\rm v}}}
\newcommand{\nusig}{\nu_{\Sigma,{\rm m}}}
\newcommand{\asig}{\alpha_{\Sigma,{\rm m}}}
\newcommand{\aave}{\alpha_{{\rm ave}}}
\newcommand{\have}{h_{{\rm ave}}}
\newcommand{\mc}{\multicolumn}

    \title[Artificial viscosity in rotating flows]{Measuring the Effects of
      Artificial Viscosity in SPH Simulations of Rotating Fluid Flows}
    \author[P. A. Taylor and
      J. C. Miller]{P. A. Taylor$^{1,2,3}$\,\thanks{E-mail:
        ptaylor@astro.ox.ac.uk (PT); jcm@astro.ox.ac.uk (JCM)} and
      J. C. Miller$^{1,4}$\,\footnotemark[1]\\$^{1}$Department of
      Physics (Astrophysics), University of Oxford, Keble Road, Oxford
      OX1 3RH, UK\\$^{2}$African Institute for Mathematical Sciences,
      6-8 Melrose Road, Muizenberg 7945, South
      Africa\\$^{3}$Department of Radiology, UMDNJ-New Jersey Medical
      School, ADMC 5, Suite 575, 30 Bergen St., Newark, NJ 07103,
      USA\\$^{4}$SISSA, International School for Advanced Studies, \&
      INFN, Via Bonomea 265, 34136 Trieste, Italy}

\begin{document}
\date{draft version}

\pagerange{\pageref{firstpage}--\pageref{lastpage}} \pubyear{2011}

\maketitle

\label{firstpage}

\begin{abstract}
A commonly cited drawback of SPH is the introduction of spurious shear
viscosity by the artificial viscosity term in situations involving
rotation. Existing approaches for quantifying its effect include
approximate analytic formulae and disc-averaged behaviour in specific
ring-spreading simulations, based on the kinematic effects produced by
the artificial viscosity. These methods have disadvantages, in that
they typically are applicable to a very small range of physical
scenarios, have a large number of simplifying assumptions, and often
are tied to specific SPH formulations which do not include corrective
(e.g., Balsara) or time-dependent artificial viscosity terms. In this
study we have developed a simple, generally applicable and practical
technique for evaluating the local effect of artificial viscosity
directly from the creation of specific entropy for each SPH
particle. This local approach is simple and quick to implement, and it
allows a detailed characterization of viscous effects as a function of
position.  Several advantages of this method are discussed, including
its ease in evaluation, its greater accuracy and its broad
applicability. In order to compare this new method with existing ones,
simple disc flow examples are used.  Even in these basic cases, the
very roughly approximate nature of the previous methods is shown.  Our
local method provides a detailed description of the effects of the
artificial viscosity throughout the disc, even for extended examples
which implement Balsara corrections.  As a further use of this
approach, explicit dependencies of the effective viscosity in terms of
SPH and flow parameters are estimated from the example cases. In an
appendix, a method for the initial placement of SPH particles is
discussed which is very effective in reducing numerical fluctuations.
\end{abstract}


\section{Introduction}
\label{sec:intro}

Continued advances in computing power and availability have led to an
increasingly important role for numerical simulations in understanding
hydrodynamical phenomena, and astrophysics provides a particular range
of interesting applications.  High-resolution, multidimensional
studies frequently include an ambitious amount of physics in order to
reproduce the dynamics of a system with sufficient accuracy. However,
numerical artefacts and limitations are necessarily present and must be
understood as fully as possible to appreciate the actual physics in
the systems being represented.

Non-physical features enter \textit{ab initio} into all hydrodynamic
simulations simply because of the discretisation of a `continuous'
fluid (which is itself discrete, of course, but on very small
scales). The two types of method developed for this task have very
different structures: Lagrangian methods follow moving fluid elements,
while Eulerian methods compute flow properties on a grid of points;
additionally, there are `hybrid' approaches, the so-called ALE
(arbitrary Lagrangian-Eulerian) schemes, e.g. the BETHE-hydro code
\citep{2008ApJS..179..209M}.  In this paper, we focus on the
Lagrangian smoothed particle hydrodynamics (SPH) method, which has
been widely used for modelling astrophysical phenomena
\citep{1977MNRAS.181..375G,1977AJ.....82.1013L}.  SPH has both
advantages and disadvantages with respect to other numerical
techniques.  One main advantage is that it automatically adapts to
following dynamic flows and arbitrary geometries, without the need for
mesh refinement or other readjustment techniques required with
Eulerian-based codes.  Furthermore, the energy equation is solved in
the local comoving frame of each fluid element, giving a direct
implementation of the first law of thermodynamics.

In modelling fluids by means of SPH particles, most SPH codes have
included an artificial viscosity term, both for making the ensemble of
discrete particles behave more like a continuum (in continuous regions
of the flow) and for handling shock discontinuities which may arise
during the course of a simulation. Methods for shock-handling with
Riemann solvers have also been implemented in SPH
\citep{2002JCoPh.179..238I,2003MNRAS.340...73C}; these approaches have
both advantages and disadvantages. Here, we focus on the artificial
viscosity methods as used in standard SPH formulations. As well as
having beneficial effects, artificial viscosity can also introduce
unwanted numerical artefacts, among which spurious shearing torques in
rotating flows are particularly troublesome (see, for example,
\citet{1994ApJ...431..754F}). Techniques such as the Balsara
correction (described in the next section) have been developed for
counteracting this. Typically, SPH prescriptions for artificial
viscosity are a composite of both bulk and shear viscosity components,
but act as a source of dissipation primarily related to shear when
modelling rotating fluid flows. It is important to have some
quantitative measure of how large the effects of this are, whether one
is dealing with problems having no intrinsic shear viscosity or
whether there is also a real, physical shear viscosity; in either
case, any numerical viscosity should be suitably negligible (unless
one is trying to use an artificial viscosity to actually model a
physical viscosity, see e.g., \citet{2010MNRAS.405.1212L}). We focus
here on two standard artificial viscosity prescriptions, with and
without the Balsara correction.

The main objective of this paper is to advocate monitoring local
entropy generation as a means for quantifying the residual effects of
artificial viscosity in regular flows not involving shocks. Our method
is completely general, but we test it out by applying it to
frequently-used idealized test problems involving isothermal discs and
compare its performance on these with that of other alternative
methods. Two main types of approach have been used previously, both of
which measure the kinematic effect of the artificial viscosity and
relate it to an effective viscosity coefficient, $\nu$. Firstly, there
are the empirical `ring-spreading' tests
\citep{1952ZNatA...7...87L,1974MNRAS.168..603L,1994ApJ...431..754F,1996MNRAS.279..402M,1999JCoAM.109..231S}.
These are based on the analytic relations for the kinematic
contribution of the artificial viscosity to disc-averaged behaviour,
with assumptions of an isothermal equation of state and a small
viscosity (so as not to generate large radial flow) that is constant
across the disc. In practice, these tests can only be performed in two
dimensions, as the thermal pressure of the fluid is zeroed to isolate
the viscosity term in the equations of motion. These methods also
require post-simulation fitting of averaged results. Secondly,
analytic approximations have been developed relating $\nu$ to SPH and
disc parameters \citep{1996MNRAS.279..402M,2010MNRAS.405.1212L}.
These relations are generally derived for specific formulations of SPH
artificial viscosity with the quadratic term set to zero.  Various
additional assumptions include having constant smoothing lengths (in
the continuum limit) and particular SPH particle kernel shapes, as
well as considerations that the viscosity is active for both receding
and approaching particles.  Both of these kinds of existing approach
involve restrictive assumptions and neither may be applied to
artificial viscosity prescriptions that include often-used
`corrective' terms, such as the Balsara method or time-dependent
approaches (discussed below), which aim to reduce `excess' shearing.

The technique presented here involves evaluating the effect of
artificial viscosity in rotating flows directly from local entropy
production (i.e. using the energy equation rather than the equation of
motion). It does not assume any particular equation of state or
rotation profile and also does not assume constancy of $\nu$ or of
smoothing lengths, or any particular artificial viscosity
prescription.  Moreover, it requires no special simulation setup and
is directly calculable from standard SPH particle quantities. Impetus
for this approach arises from the nature of the SPH method itself and
its direct application of the first law of thermodynamics, noting that
a clear understanding of entropy production is of great importance in
numerical simulations. The technique is simple and quick to implement,
and it allows a detailed characterization of the effects of the
artificial viscosity as a function of position.  Several advantages of
this method over existing ones are discussed, including its ease in
evaluation, its direct interpretation, its greater accuracy and its
broad applicability, e.g., to arbitary rotation profiles and equations
of state.

For all of the simulations described here, we have utilised the
public-release version of the Gadget-2 code
\citep{2005MNRAS.364.1105S}.  While several artificial viscosity
formulations of varying sophistication have been developed over the
years, in this paper we focus mainly on the {\it method} of estimating
the effects of this viscosity and therefore utilize only the
relatively simple form that is standard in Gadget-2 (with and without
the Balsara treatment, described below). However, the method itself is
independent of viscosity prescription, and our aim is to apply it to a
comparison of several other formulations in a forthcoming paper.
Here, we begin by briefly reviewing in \S\ref{sec:newtonhydro} the
form of SPH equations used and considerations concerning the
artificial viscosity; in \S\ref{sec:rotflowartvisc}, the
ring-spreading test is described; in \S\ref{sec:motiv}, both the
motivation for and the general formulation of our local criterion for
measuring the effective viscosity in rotating flows are described; in
\S\ref{sec:general_spec}, example applications of our method in thin
discs are given, and various simulations performed, for which the
results of the local measures of viscosity are compared with those of
the ring-spreading tests and existing analytic formulations; these
results are then discussed in \S\ref{sec:discuss}.  Additionally, in
Appendix \ref{appendixa}, we discuss the SPH particle setup used for
our initial conditions, which is presented as a simple improvement to
aid in reducing numerical effects and increasing the efficiency of
convergence.

\section{SPH, using Gadget-2}
\label{sec:newtonhydro}

In SPH a continuous fluid is sampled at a finite number of points, and
discretised versions of the standard hydrodynamic equations of
continuity, momentum (Euler) and the first law of thermodynamics
(energy equation) are evolved, given respectively in their simplest
form as:
\begin{eqnarray}
0&=&\frac{{\rm d}}{{\rm d}t}\rho + \rho\nabla \cdot {\bf v} \label{eqcont}\\
0&=& \frac{{\rm d}}{{\rm d}t} {\bf v}+\frac{1}{\rho}\nabla P \label{eqcom}\\
0&=&\frac{{\rm d}}{{\rm d}t}\epsilon - \frac{P}{\rho^2}\frac{{\rm d}}{{\rm d}t}\,\rho\, ,\label{eqener}
\end{eqnarray}
written with Lagrangian derivatives,
${\rm d}/{\rm d}t=\partial/\partial t+{\bf v}\cdot\nabla$,
and with mass density, $\rho$; velocity, ${\bf v}$; thermal pressure,
$P$; and specific internal energy, $\epsilon$.  The exact form of SPH
implementations has continued to be refined over the decades; in this
work, we have used the publicly-available SPH code, Gadget-2, the
formulation of which is derived from a fluid Lagrangian and,
importantly, directly conserves linear and angular momentum, entropy
and energy \citep{2002MNRAS.333..649S}.  The Gadget-2 SPH equations are
presented here, with emphasis on the necessity for including extra,
purely numerical terms in the implementation of the continuum
equations above.  For further details, we refer the reader to much
more comprehensive discussions elsewhere, e.g.
\citep{2005RPPh...68.1703M,2010arXiv1012.1885P}.

The fluid is taken to have a polytropic equation of state (EOS),
\begin{equation} \label{polytr_eos}
P=K(s)\rho^{\gamma}=(\gamma-1) \rho \epsilon\,,
\end{equation} 
where $s$ is the specific entropy.  Following standard usage (e.g.,
\citet{1959flme.book.....L}), a polytropic gas is defined as having
constant specific heats, and therefore the related adiabatic index,
$\gamma$, is constant. While some authors restrict $K$ to being
constant as well (and refer to it as the polytropic constant), here it
is allowed to vary, and in the following $K$ is referred to as the
entropic function.  In Gadget-2, this $K$ is evolved rather than
$\epsilon$, giving the strict conservation properties of both energy
and entropy mentioned above \citep{2002MNRAS.333..649S}. In the
following, the subscripts $\{a, b\}$ will be used only to distinguish
individual fluid elements, represented in SPH as finite volume
pseudo-particles, and not to signify vector components.

First, the mass density at the location of a given particle labelled
as `$a$' and with spatial coordinate vector, ${\bf r}_a$, is
calculated at each timestep by direct summation using
\begin{equation}\label{sph_cont}
\rho_a=\displaystyle\sum_{b}^{N_{\rm n}} m_b W_{a}\,.
\end{equation}
The summation is performed over ${N_{\rm n}}$ `nearest neighbour'
particles labelled as `$b$' (typically $\approx 50$
\citep{1993MNRAS.265..271N}), each with mass, $m_b$, and contained
within a radius given by the smoothing length, $h_{a}$.  The kernel,
$W_{a}= W(|{\bf r}_{ab}|,h_{a})$, is strongly peaked and
differentiable (a cubic spline, in Gadget-2) and with $|{\bf
  r}_{ab}|=|{\bf r}_{a} -{\bf r}_{b}| < h_{a}$.  The smoothing length
is calculated at each timestep such that the kernel volume of a
particle contains a constant mass (for numerical stability), according
to the following relation:
\begin{equation}\label{sph_hsml}
\frac{4\pi}{3}h^3_{a}\rho_a = N_{\rm n} \bar{m} ={\rm const.},
\end{equation}
where $\bar{m}$ is the average particle mass. Within the fluid, the
resolution is essentially determined by the magnitude of $h_{a}$, over
which discontinuities are smoothed.

To evolve the particle velocity, the non-viscous Euler equation,
including a gravitational potential, $\phi$, is represented as:
\begin{equation}\label{sph_euler}
\left(\frac{{\rm d}{\bf v}_a}{{\rm d}t}\right)_{\rm nv} = - \displaystyle\sum_{b}^{N_{\rm n}}
m_b \left( f_b \frac{P_b}{\rho^2_b}\nabla W_{b} + f_a
\frac{P_a}{\rho^2_a}\nabla W_{a} \right) +\nabla\phi_a,
\end{equation}
where
\begin{equation}
f_a = \left(1 +\frac{h_{a}}{3 \rho_a}\frac{\partial \rho_a}{\partial
  h_{a}} \right)^{-1}. \nonumber
\end{equation}
The factors, $f_a$, arise from the Lagrangian derivation and the set
of constraints on coordinates provided by Eq.~\ref{sph_hsml}; they
account directly for the variation in smoothing lengths in the system
\citep{2002MNRAS.333..649S}.  Self-gravity among particles is
calculated efficiently using a tree algorithm (a hierarchical
multipole expansion) \citep{1986Natur.324..446B}.  Additionally, in
some cases we include the acceleration due to a (Newtonian) central
object located at the origin of the simulation's coordinates,
$\nabla\phi_{a}=-GM_{{\rm C}} \hat{ \bf r}_a/r_a^2$.

Gadget-2 uses an artificial viscosity of the form
\begin{equation}\label{visc_sph}
\Pi_{ab}=\left\{
\begin{tabular}{cl}
$-\av(\bar{c}_{ab}w_{ab} -  3w^2_{ab})/2\bar{\rho}_{ab}$ &\mbox{for 
${\bf v}_{ab}\cdot{\bf r}_{ab}<0$},\\[0.2cm] 
$0$ & \mbox{for ${\bf v}_{ab}\cdot{\bf r}_{ab}\geq 0$},
\end{tabular}\right.
\end{equation}
with ${\bf v}_{ab}={\bf v}_{a}-{\bf v}_{b}$ and $w_{ab}=({\bf
  v}_{ab}\cdot{\bf r}_{ab})/|{\bf r}_{ab}|$ being the relative
velocity between particles in vector and scalar form,
respectively; $\bar{\rho}_{ab}=(\rho_a+\rho_b)/2$, the average
density; $c_a\,=\,(\gamma P_a/\rho_a)^{1/2}$, the local soundspeed,
with average $\bar{c}_{ab}=(c_a+c_b)/2$; and $\av\approx1$, a free
parameter for the strength of the viscosity.  In practice, $\Pi_{ab}$
acts similarly to a pressure term in the Euler equation, and the
related acceleration is:
\begin{equation}\label{sph_eul_visc}
\left(\frac{{\rm d}{\bf v}_a}{{\rm d}t}\right)_{{\rm v}}\,=\,
- \displaystyle\sum_{b}^{N_{\rm n}} m_b \Pi_{ab}{\bf \nabla} \bar{W}_{ab}\,,
\end{equation}
where $\bar{W}_{ab}\, = (W_{a}+ W_{b})/2$.

The $\Pi_{ab}$ formulation possesses two terms to mimic
naturally-occuring, dissipative processes.  The first (linear in
$w_{ab}$) provides both the bulk and shear viscosity of the converging
particles, and the second (quadratic in $w_{ab}$) functions as a von
Neumann-Richtmyer artificial viscosity for shock handling and for
spreading shock discontinuities over the smoothing length of
supersonic particles.  The quadratric term also prevents interparticle
penetration. It should be noted that this single artificial viscosity
provides both shear and bulk viscosities, which would be given by
separate terms in the physical, Navier-Stokes description (direct
implementations of the latter are discussed briefly below).  In many
SPH prescriptions the linear and quadratic terms are scaled by
separate free parameters, $\av$ and $\bv$, respectively, each with a
range of `typical' values, but we here follow the practice of setting
$\beta=3\alpha$ (unless otherwise stated).  An in-depth discussion of
the relative scaling of the linear and quadratic terms is provided by
\citep{1997JCoPh.136..298M}.

The entropic function, $K$ (defined in Eq.~\ref{polytr_eos}),
changes, in general, in the presence of viscosity:
\begin{equation}\label{sph_energ}
\frac{{\rm d}K_a}{{\rm d}t}=\frac{1}{2} \left(\frac{\gamma -
  1}{\rho_a^{\gamma-1}}\right) \displaystyle\sum_{b}^{N_{\rm n}}
m_b \Pi_{ab} {\bf v}_{ab} \cdot {\bf \nabla} \bar{W}_{ab}\,.
\end{equation}
Cooling in optically thin material may be included in a
straightforward manner by subtracting the appropriate entropy loss at
the end of a timestep. The inclusion of accurate radiative transport
in optically thick material is a separate (and highly non-trivial)
task, which we do not consider here.

Finally, to limit spurious angular momentum transfer which arises in
shear flow due to the form of the artificial viscosity, the simple
Balsara correction is commonly utilised
\citep{1995JCoPh.121..357B,1996MNRAS.278.1005S}. The aim of this
factor is to remove the effect of the artificial viscosity when there
is a pure shearing motion and to have it acting only when there is a
compression.  An estimate of the relative amount of local shear is
made from the curl and divergence of the particle velocities:
\begin{equation}\label{balsara}
g_a=\frac{| \nabla\cdot {\bf v}|_a}{| \nabla \times
  {\bf v}|_a+| \nabla \cdot {\bf v}|_a +
  (0.0001\,c_a/h_{a}) }\,,
\end{equation}
where the third term in the denominator prevents the quantity from
diverging.  The average value of this quantity for two interacting
particles is used as a multiplicative factor in front of $\Pi_{ab}$,
producing the modified form $\Pi^{\prime}_{ab} = \Pi_{ab}\,
(g_a+g_b)/2$, which is then used instead of $\Pi_{ab}$ in
Eqs.~\ref{sph_eul_visc} and \ref{sph_energ}.  In the following, we
examine the effect of the artificial viscosity both with and without
the Balsara correction (the former case being the default).

Eqs.~\ref{sph_cont}, \ref{sph_euler} and \ref{sph_energ},
together with the EOS and the smoothing length condition in
Eq.~\ref{sph_hsml}, form a complete set of equations for evolving
the physical, hydrodynamic quantities in numerical simulations.
However, as part of the process of discretisation, a number of
additional variables and parameters have been included ($h$, $W$,
etc.).  While these are used to model realistic properties of the
discretised system, such as continuity and shocks, they also introduce
purely numerical features by controlling resolution, the `spread' of
shocks and the stability of interpolations and summations.  In the
past decades, much work has been done to determine reliable values and
forms for these parameters and to reduce associated numerical
artefacts. Generally, a number of standard problems, in various
dimensions and geometries, are used to test the behaviour of a given
code and the associated parameters.  

Another purely numerical consideration in using SPH is the initial
placement of the particles representing points in the flow
\citep{2002ApJ...569..501I,2006MNRAS.365..199M}.  For instance,
regularities in particle configurations may lead to the propagation of
artificial structures in the simulation along preferred directions,
and even randomised placements may produce numerical features, e.g.,
due to local overdensities.  For each simulation presented here, the
initial particle setup was created using an algorithm which maps a
non-regular but constant number density profile of (equal-mass)
particles into the arbitrary number-density (and trivially,
mass-density) profile of the desired model.  This method has also been
applied in other simulations \citep{2011MNRAS.410.2385T} and is
described in Appendix~\ref{appendixa}.

\section{Existing (kinematic) approximations of $\nu$}
\label{sec:rotflowartvisc}
The standard ring-spreading test starts from an initial
$\delta$-function ring of matter in circular Keplerian motion around a
gravitating point mass
\citep{1952ZNatA...7...87L,1974MNRAS.168..603L}. If shear viscosity is
present, this ring will proceed to spread out into a disc, with the
rate of spreading dependent on the magnitude of the viscosity.  The
matter is taken here to behave isothermally, with the temperature and
viscosity being sufficiently small so that the spreading is slow and
the thin disc is essentially Keplerian (with its height, $H$, being
small compared with the radial coordinate, $r$, at any point)
\citep{1981ARA&A..19..137P}. With certain additional simplifying
assumptions (negligible self-gravity for the disc material and
constant viscosity coefficient, $\nu$), an analytic solution can be
obtained for the ring-spreading
\citep{1994ApJ...431..754F,1996MNRAS.279..402M,1999JCoAM.109..231S}.

The exact analytic results provide a means for comparison for an SPH
simulation of an equivalent configuration with no viscosity apart from
that of the numerical scheme, $\nu_{\rm SPH}$. From the analytic
results, an effective value for the shear component of the artificial
viscosity can then be estimated, if one assumes that $\nu_{\rm SPH}$
is roughly constant throughout the disc (a somewhat doubtful
assumption) and that the corresponding bulk component has negligible
effect (a very good assumption under the circumstances envisaged).
The value obtained can then be translated into an effective
Shakura-Sunyaev $\alpha$ coefficient [25], which represents the
viscosity according to
\begin{equation}\label{shaksuny}
\nu=\alpha\,c_{\rm{s}}\,H,
\end{equation}
where $H$ is the disc height, and $c_{\rm{s}}$ is the sound speed.

The analytic solution gives an expression for the surface density of
the ring, $\Sigma_{\rm R}$, as a function of position and time. For
the conditions mentioned above, the radial velocity $v_r$ (positive
for `outer' material and negative for `inner' material) is always
small, with $v_r \sim \nu/r \ll v_\phi$ by scaling arguments
\citep{2002apa..book.....F}. The surface density is then given by
\begin{equation}\label{sigmathin}
\Sigma_{{\rm R}}(r, \tau) = \frac{M_{{\rm D}}}{\pi \tau
  r^{1/4}}\,\exp\left(\frac{-1-\textit{r}^2}{\tau}\right)\,I_{1/4}
\left(\frac{2\textit{r}}{\tau}\right),
\end{equation}
where $r$ is the radius in the equatorial plane (with $r = 1$
initially); $\tau$ is a time parameter ($\tau = 12 \nu t$); $\nu$ is
the kinematic shear viscosity coefficient; $M_{\rm D}$ is the disc
mass (taken as the mass unit, i.e. $M_{\rm D} = 1$), and $I_{1/4}$ is a
modified Bessel function. A detailed derivation of this formula is
given by, e.g., \citet{2002apa..book.....F}.

In our work reported below, the corresponding numerical simulations
begin a short time after the axisymmetric ring has spread to a finite
thickness (at $\tau_0 = 0.01$) using the analytic formula
Eq.~\ref{sigmathin} to provide the initial surface density profile.
This same formula is also used for matching against the simulation
properties at subsequent times (doing so has been justified in
previous numerical studies by the general proximity of the surface
density evolution to this expression).  Self-gravity of the SPH
particles is included in simulations, although it is assumed that the
acceleration due to the central point mass dominates (we take $M_{{\rm
    C}}/M_{{\rm D}}=1000$). SPH particles are removed from the system
at an inner boundary with small radius, $r = 0.1$.

As mentioned above, in addition to the ring-spreading test there are
also some approximate analytic formulations for estimating the
effective shear viscosity of the disc from kinematic considerations:
that of \citet{1996MNRAS.279..402M} is
\begin{equation}
  \nu_{\rm M} = \alpha_{\rm v} c_{\rm s} h/16\,,
\end{equation}
while that of \citet{2010MNRAS.405.1212L} is
\begin{equation}
  \nu_{\rm LP} = \alpha_{\rm v} c_{\rm s} h/20\,.
\end{equation}
These relations are generated essentially by reversing the summation
form of SPH equations which involve artificial viscosity back to
continuum limits.  Further details are given in the respective
papers. Note that the Gadget-2 definition of smoothing length has been
used in each case.

\section{Estimating local $\nu$ from entropy or viscous 
energy production}
\label{sec:motiv}
\subsection{Motivation from fluid equations}

An artificial viscosity term ($\propto \Pi_{ab}$) appears in both the
momentum and energy hydrodynamic equations. Each of these roles
affects the evolution and structure of the system, as the trajectories
of fluid elements are modified and as kinetic energy is transformed
into heat.  The standard form of the (isothermal) ring-spreading test
has been used essentially to analyse the effective viscosity via the
Euler equation, with the spreading being caused by the combined action
of the viscous and several other additive terms\footnote{In practice,
  ring-spreading is performed in 2D only, with the thermal pressure
  terms zeroed in a `cold' disc approximation to isolate the viscous
  term, so that the 3D behaviour of the artificial viscosity is
  typically not measured.}, but it seems better to use, instead, the
energy equation, where the effect of viscosity in generating entropy
can be clearly singled out.

Internal energy in fluid systems can be produced both by viscosity and
by non-viscous compression.  This is expressed in the first law of
thermodynamics, as applied to a co-moving fluid element with unit
mass:
\begin{equation}\label{firsttherm}
  d\epsilon = T\,ds - P\,d(1/\rho)\,,
\end{equation}
where $T$ is the temperature and $s$ is the specific entropy. The
specific internal energy produced by viscosity, $d\epsilon_{\rm v}$,
corresponds to the first term on the right hand side, and its time
derivative is simply
\begin{equation}\label{visc_en_entr}
  \dot{\epsilon}_{\rm v} \equiv T\dot{s}\,.
\end{equation}
This quantity can be calculated directly from the SPH expressions in
all cases, whatever the flow geometry or equation of state, giving a
direct local measure of the effects of viscosity. The only caveat is
that there should be no other sources of entropy production to
contaminate the interpretation.

For the particular case of the (widely used) class of polytropic
equations of state implemented by Gadget-2, the derivative of
Eq.~\ref{polytr_eos} gives the following expression for $d\epsilon$:
\begin{eqnarray} \label{polytr_first_law}
   d\epsilon&=&
   \frac{\rho^{\gamma-1}}{\gamma-1}dK+K\rho^{\gamma-2}d\rho\,,\nonumber\\
   &=& \frac{\rho^{\gamma-1}}{\gamma-1}dK-P\,d\left(\frac{1}{\rho}\right)\,,
\end{eqnarray}
where the relations $P = K\rho^{\gamma}$ and $d(1/\rho) =
-d\rho/\rho^2$ have been used in arriving at the form of the second
line.  Matching terms with the first law in Eq.~\ref{firsttherm} and
referring to Eq.~\ref{visc_en_entr} leads to the following set of
formulae linking $\dot{K}$, $\dot{s}$ and $\dot{\epsilon}_v$:
\begin{equation}\label{visc_entr_entr}
   \dot{K} = \frac{(\gamma-1)T}{\rho^{\gamma-1}}\,\dot{s}=
   \frac{(\gamma-1)}{\rho^{\gamma -1}}\, \dot{\epsilon}_v  \,.
\end{equation}
Within Gadget-2, computation of the evolution of $\epsilon_{\rm v}$ is very
similar to that for the entropic function, $K$ (cf. Eqs.~\ref{sph_energ},
\ref{visc_en_entr} and \ref{visc_entr_entr}):
\begin{equation}\label{sph_evisc}
\frac{{\rm d}\epsilon_{{\rm v}a}}{{\rm d}t} = 
\left( \frac{\rho_a^{\gamma-1}}{\gamma - 1} \right)
\frac{{\rm d}K_{a}}{{\rm d}t}=
\frac{1}{2} \displaystyle\sum_{b}^{N_{\rm n}} m_b \Pi_{ab}
{\bf v}_{ab} \cdot {\bf \nabla} \bar{W}_{ab}\,,
\end{equation}
and this can be evaluated straightforwardly for each fluid element.
Note that in the further specialisation to an isothermal equation of
state with $\gamma = 1$ (which we will be using below for making
comparison with the ring-spreading test), then the entropic function,
$K$, becomes a constant, but the true entropy, $s$, continues to
change under the action of viscosity, as does $\epsilon_{\rm v}$.

Importantly, $\dot{\epsilon}_{\rm v}$ is determined from the
contribution of only a single term which contains $\Pi_{ab}$. In
quantifying the effect of any SPH artificial viscosity prescription on
a disc system, this isolation greatly simplifies analysis (and
interpretation). In general, entropy can be produced by shearing,
normal compression and shock compression, the totality of which are
accounted for when measuring $\dot{\epsilon}_{\rm v}$, due to its
general relation to specific entropy. In contrast, only the shearing
is important under the assumptions of the ring-spreading test (or the
analytic approximations).

\subsection{Local viscous heating of fluid elements}
\label{sec:general_der}

Monitoring $\dot{\epsilon}_{\rm v}$ already provides a suitable way of
demonstrating the local effects of artificial viscosity, and one could
stop at that point. However, if one wants to calculate an effective
shear viscosity coefficient (for circumstances involving shear and no
significant compression), one needs to write an expression for
$\dot{\epsilon}_{\rm v}$ in terms of $\nu$ and the shearing velocity
field (see, e.g., Appendix B of \cite{1978trs..book.....T}) and then
to invert it to give an expression for $\nu$.  For general planar
rotational motion, one finds the following expression for specific
internal energy creation \citep{1997MNRAS.287..165S}:
\begin{equation}\label{szmil}
  \dot{\epsilon}_{\rm v} = \frac{4}{3}\nu \left[ 
\left(\frac{\partial v_r}{\partial r} \right)^2+
\frac{ v_r^2}{r^2}-
\frac{\partial v_r}{\partial r}\frac{ v_r}{ r}
\right] + \nu \left(r
  \frac{\partial\Omega}{\partial r} \right)^2\,,
\end{equation}
using cylindrical polar coordinates ($r, \phi, z$) here and after.
This form may be inverted easily for an expression of the effective
viscosity:
\begin{equation}\label{nu_szmil}
  \nu(r) = \frac{3\dot{\epsilon}_{\rm v}}{4} \left[ 
\left(\frac{\partial v_r}{\partial r} \right)^2+
\frac{v_r^2}{ r^2}-
\frac{\partial v_r}{\partial r}\frac{ v_r}{ r} + \frac{3}{4} \left(r
  \frac{\partial\Omega}{\partial r} \right)^2\right]^{-1}.
\end{equation}
Each term on the right hand side is composed of variables which are
directly obtainable from the SPH particles. As discussed above, the
viscous heating is a direct measure of entropy creation due to
viscosity, and the gradient terms may be estimated, e.g., within SPH
particle kernels or from averaged ring values.

In this study, we describe and utilise an expression which includes
solely the term involving $\Omega$. This facilitates comparisons with
existing approximations and, as shown below, leads to particularly
simple expressions of $\nu(r)$ (though, it must be emphasised, that
the preceding expression is by no means prohibitively complicated).

The surface area of a narrow, axisymmetric ring of width, $\Delta r$
(centred around $r$), is $2\pi r \Delta r$, and the mass of the ring
is therefore $\Delta m = 2 \pi r\Sigma\Delta r $.  Taking the matter
to be moving on basically circular orbits, the viscous torque exerted
on the ring can be written as \citep{2002apa..book.....F}:
\begin{equation} \label{ring_torq}
G(r)=\frac{d\Omega}{dr} 2 \pi r^3 \nu \Sigma \,.
\end{equation}
The work done by this torque leads to local heating in the rotating
flow. The specific heating rate due to viscosity is
\begin{eqnarray}
  \dot{\epsilon}_{{\rm v}}(r) &=& \frac{d\Omega}{dr} \frac{G(r)\,\Delta r}{\Delta m} = \left(\frac{d\Omega}{dr}\right)^2 \frac{2\pi r^3\nu\Sigma\, \Delta r\,}{\Delta m},\nonumber \\
  &=&\nu \left(r\frac{d\Omega}{dr}\right)^2\,,
\end{eqnarray}
which is recognized as the final term in Eq.~\ref{szmil}.

After trivial rearrangement, the expression for the viscosity is given
in terms of easily known quantities as:
\begin{equation} \label{general_visc}
\nu(r) =  \left(r\frac{d\Omega}{dr}\right)^{-2}   \dot{\epsilon}_{{\rm v}}\,.
\end{equation}
Compared to existing methods for evaluating effective viscosity in
rotating flows, this expression is particularly general, having made
no assumptions of rotation profile, equation of state, constancy of
$\nu$ or artificial viscosity prescription.  Moreover, as a function
of radius, this expression provides a more detailed and useful
description of viscous behaviour throughout the simulation.  As noted
above, more general viscous stresses may be applied as well, such as
that of Eq.~\ref{szmil}.  In the following examples, we implement
Eq.~\ref{general_visc} with different disc stuctures and using
different artificial viscosity prescriptions.

\section{Examples and Comparisons}
\label{sec:general_spec}
\subsection{Local heating applied to thin discs}
In the examples below, we apply the expression for the effective local
viscosity in Eq.~\ref{general_visc} to test scenarios of
ring-spreading and thin disc cases.  Using assumptions of axisymmetry
and circular, Keplerian orbits, the viscosity expression is
\begin{equation}\label{newnucrit}
\nu(r) = \frac{4}{9} \frac{\dot{\epsilon}_{{\rm v}}}{\Omega_{{\rm K}}^2}\,.
\end{equation}
Presuming vertical equilibrium with some thin disc scale height, $H$,
the Shakura-Sunyaev $\alpha$ is then given as
\begin{equation}\label{newvisccrit1}
  \alpha(r) =  \frac{4}{9}
  \frac{\dot{\epsilon}_{{\rm v}}}{\Omega_{{\rm K}}^2c_{\rm s}H}\,.
\end{equation}

In the case of polytropic, nonisothermal equations of state, these
expressions for viscosity and the Shakura-Sunyaev $\alpha$ may be
written in terms of the entropic function as:
\begin{eqnarray}
\nu(r) &=& \frac{4}{9}\label{newnucrit2}
\frac{\rho^{\gamma-1}\dot{K}}{(\gamma-1)\Omega_{{\rm K}}^2},\\
\alpha(r) &=& \frac{4}{9}\label{newvisccrit2}
\frac{\rho^{\gamma-1}\dot{K}}{(\gamma-1)\Omega_{{\rm K}}^2c_{\rm s}H}\,,
\end{eqnarray}
respectively. These last forms are particularly relevant to Gadget-2,
where the entropic function is directly evolved, and they are used in
obtaining the results presented below.

Finally, it is interesting to note that, for the specific case of an isothermal
EOS, Eq.~\ref{newvisccrit1} may be expressed neatly as:
\begin{equation}\label{newvisccrit}
  \alpha(r) = \frac{4}{9} \frac{\dot{\epsilon}_{{\rm v}}}{
    \Omega_{{\rm K}} \,c_{{\rm s}}^2 }\propto
  \frac{t_{{\rm dyn}}}{t_{{\rm th,v}} }\,,
\end{equation}
using the expression for the scale height given in Eq.~\ref{thinstruc},
below. Physically, in this latter form, $\alpha$ directly relates the
local dynamical timescale ($t_{{\rm dyn}}=\Omega^{-1}$) and the
viscous thermal timescale ($t_{{\rm th,v}} \propto c_{{\rm
    s}}^2/\dot{\epsilon}_{{\rm v}}$), as it should from accretion disc
theory (e.g., \citet{1981ARA&A..19..137P}).

\subsection{Thin Disc Structure}
\label{sec:thindiscstr}
We now briefly describe the structure equations for the two varieties
of thin isothermal discs used in this study (in dimensionless
units).  Each disc is taken to be in vertical,
hydrostatic equilibrium with respect to the gradient of the
gravitational potential of the massive central object. The density
distribution, scale height and central density (in the equatorial
plane) are, respectively:
\begin{eqnarray}\label{thinstruc}
\rho(r, z) &=&\rho_{\rm{c}}(r)\,\exp\left[-\frac{1}{2}\left(\frac{z}{H(r)}\right)^2\right],\nonumber \\
 H(r) &=&\left[\frac{r^3c^2_{\rm{s}}}{GM_{\rm{C}}}\right]^{1/2}= \frac{\,c_{\rm{s}}}{\Omega_{\rm{K}}}\,,\nonumber 
  \\ 
\rho_{\rm{c}}(r) &=&\frac{\Sigma(r)}{[2\pi]^{1/2}H(r)}\,. 
\end{eqnarray}
Since this Gaussian density profile only asymptotically approaches
zero in the vertical direction, an arbitrary disc height for the
boundary of SPH particles above/below the equatorial plane at any
radius must be designated, and we set this to be three times the scale
height in order to incorporate $>99\%$ of the mass of the theoretical
distribution into the simulation (though, the non-arbitrary scale
height, $H$, is used everywhere else in the analytic relations and
discussion). 

The vertical velocity is taken to be uniformly zero, and the profile
of the azimuthal velocity is very nearly Keplerian, i.e., $v_{\phi}(r,
z)\approx v_{\rm{K}}(r)$. The global sound speed, $c_{\rm{s}}$, is a
parameter restricted in accordance with the thinness condition, $H\ll
r\Rightarrow c^2_{{\rm s}}\ll GM_{{\rm C}}/r$, or $v^2_{{\rm
    K}}/c^2_{{\rm s}}\equiv\mu^2\,\gg\,1$, where $\mu$ is essentially
the Mach number of the (very supersonic) azimuthal flow.  Using both
the height from Eq.~\ref{thinstruc} and the definition of the
viscosity parametrisation in Eq.~\ref{shaksuny}, the global disc
viscosity can be related to the Shakura-Sunyaev $\alpha$ as a function
of $\mu$ and the Keplerian specific angular momentum, $j_{{\rm K}}$:
\begin{equation}\label{visc_mach}
\nu=  \alpha\,\frac{r v_{\rm K}}{\mu^2}=
\alpha\, \frac{j_{\rm K}}{\mu^2}  \,.
\end{equation}

Finally, the initial surface density of the flows must be supplied for
use in Eqs.~\ref{thinstruc}.  The formulation of the spreading ring,
$\Sigma_{{\rm R}}$, has been given in Eq.~\ref{sigmathin}.  We have
also calculated a more disc-like system (of similar mass and surface
density), with radial span at $t=0$ of $(r_{{\rm min}},\,r_{{\rm
    max}})=(0.7,\,1.4)$.  Here, the arbitrary surface density profile
has been chosen to be flat in this region initially, $\Sigma_{{\rm
    F}}(r,\,0)=\Sigma_{{\rm R}}(1,\,\tau=0.01)/2$, and the resulting
disc mass was $\approx2$.  This is considered solely as a separate
test case and not as a representation of a specific, physical system
(e.g., the initial boundaries are too discontinuous for an equilibrium
system).

\subsection{Global and local viscosity estimates} 
\label{sec:SPHvisc_ring}

Here, we first present results for ring-spreading tests with the
Balsara correction included for various values of the parameter,
$\mu_1\equiv\mu(r=1)$, where $r=1$ is the radius of the initial
midpoint of the ring (and using units in which $GM_{\rm
  C}=0.01$). First, the viscous properties are determined
kinematically by using the global evolution of the surface density and
fits to the theoretical progression given by the analytic formulation.
Then, local values of $\nu$ and $\alpha$ are calculated from
$\dot{\epsilon}_{\rm v}$ using
Eqs.~\ref{newnucrit}-\ref{newvisccrit1}.  Unless stated otherwise,
$\av=0.8$ in all of the presented simulations (and, in the notation of
\S\ref{sec:newtonhydro}, the quadratic term coefficient, $\bv=2.4$).

\begin{figure}
    \begin{tabular}{c}
      \includegraphics[width=2.2in]{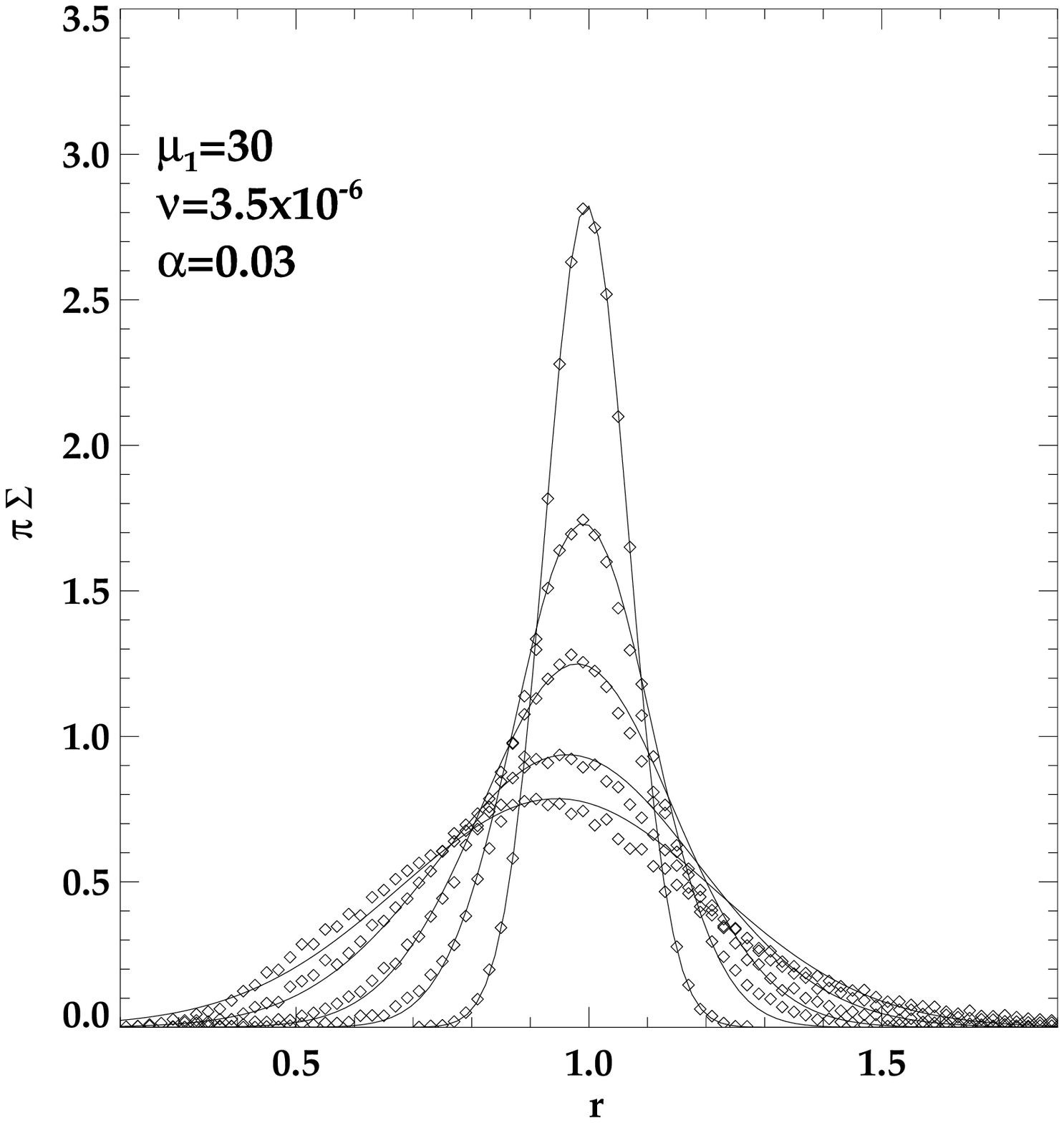}\\
      \includegraphics[width=2.2in]{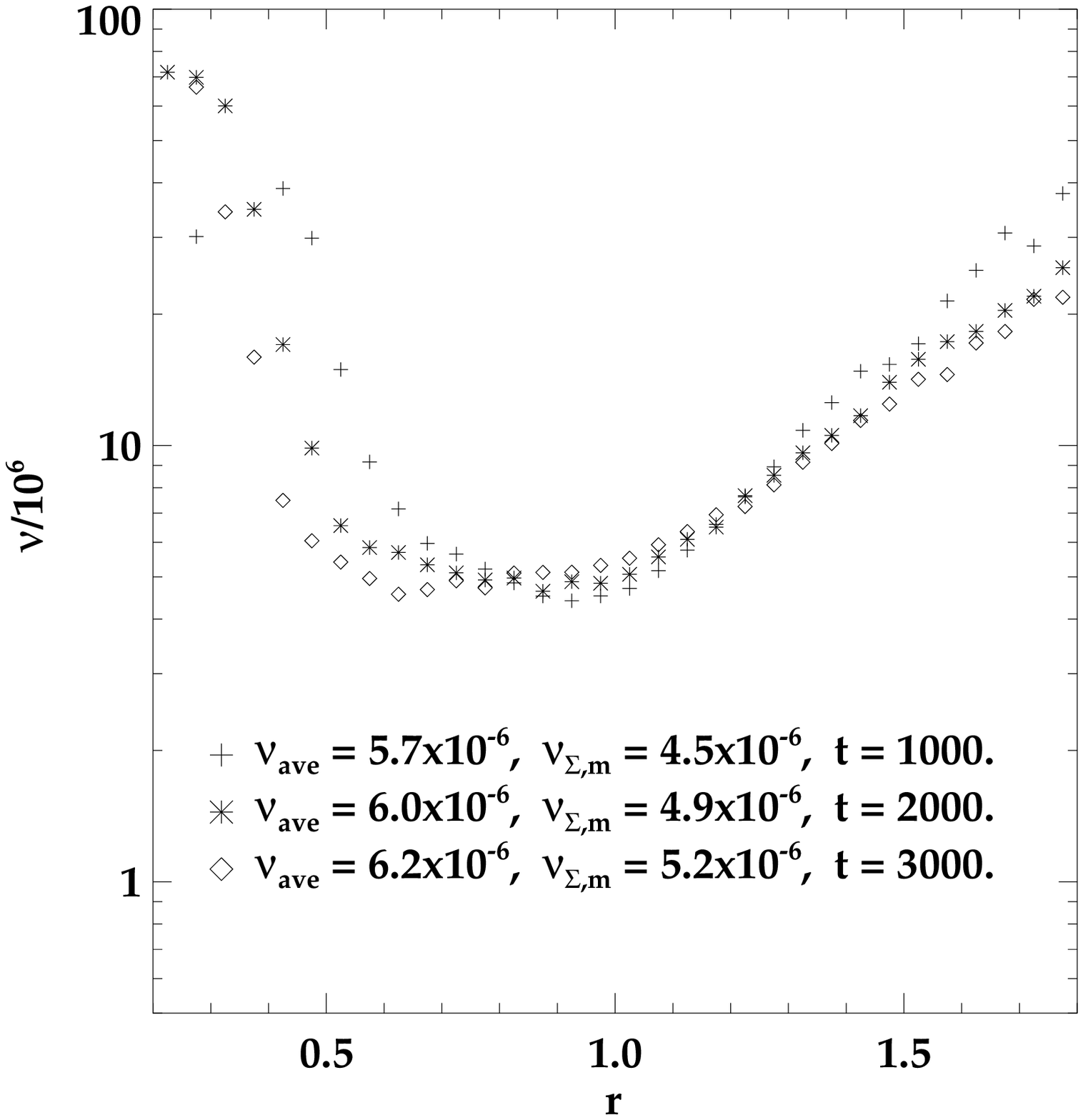}\\
      \includegraphics[width=2.2in]{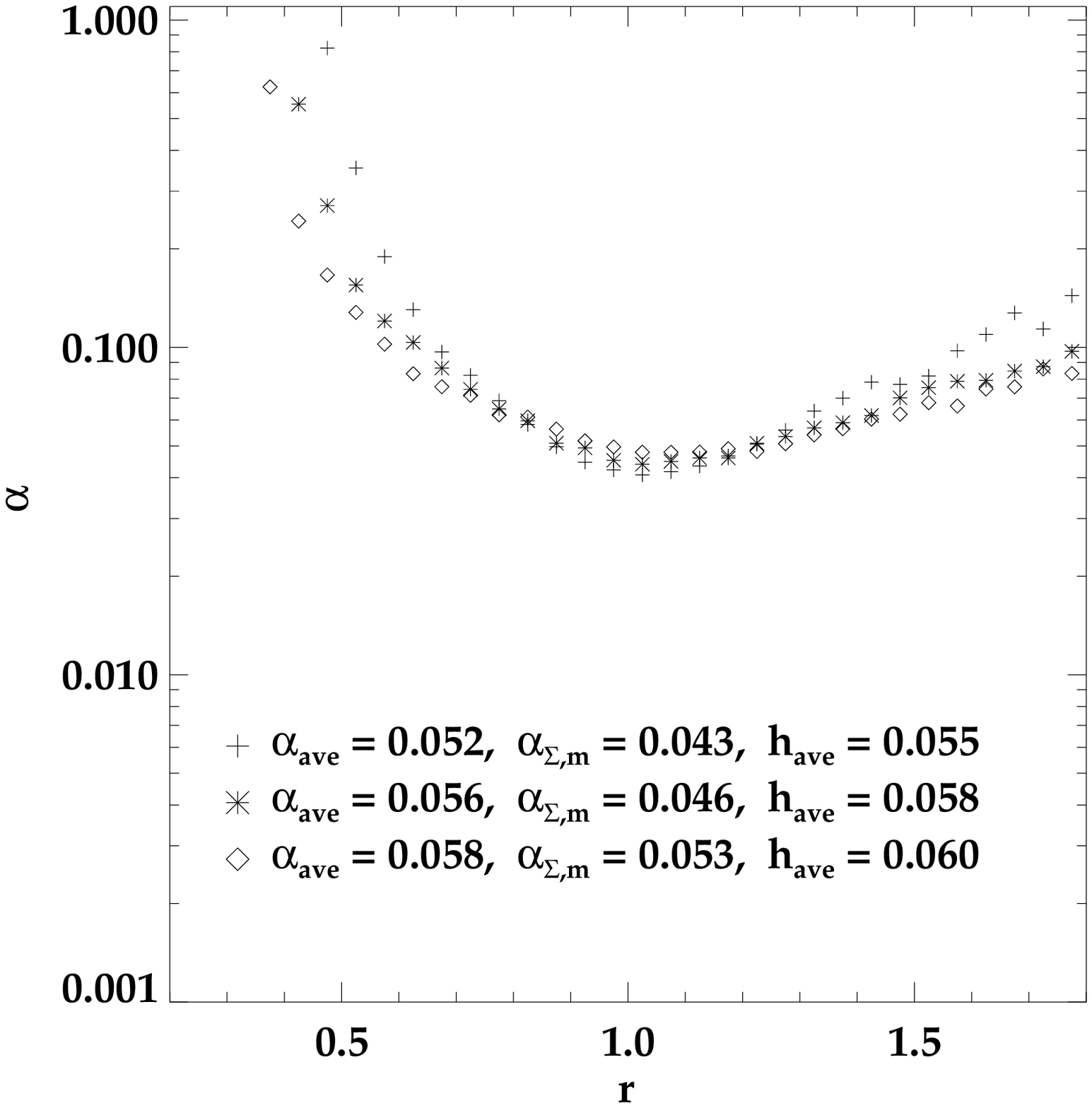}            
  \end{tabular}
  \caption{\label{fig:ringsp_cs30} (Here and following:) Top: standard
    $\nu$ and $\alpha$ estimation from $\Sigma_{{\rm R}}$ evolution
    (`$\diamond$') with analytic solution (solid line), shown at $t=0$,
    400, 1000, 2000 and 3000.  Middle: local $\nu$ (average and value
    at maximum $\Sigma$) from Eq.~\ref{newnucrit}.
    Bottom: local $\alpha$ from $\nu$ values in middle panel, using
    Eq.~\ref{newvisccrit1}, and average smoothing length,
    $\have$. $N$=50k.}
\end{figure}

The top panel of Fig.~\ref{fig:ringsp_cs30} shows the $\Sigma_{{\rm
    R}}$ time evolution of a ring with $\mu_1=30$ at a resolution of
$N=5\times10^4$ (binned values plotted as diamonds).  The value of
$\nu=3.5\times10^{-6}$ can then be inferred by matching with analytic
fits given by Eq.~\ref{sigmathin}, which are also shown (solid
lines). Use of Eq.~\ref{visc_mach} then easily yields values of
$\alpha\approx0.03$ at $r\approx1$. The profile from the simulation
roughly matches the analytic expectation, and while fitting the
curves, more weight was given to the locations of peak surface
density, with the inner and outer margins being slightly over- and
under-dense, respectively.  It is striking that there is non-neglible
effective shear viscosity present in these simulations, even with the
Balsara correction.

The local estimates of viscosity parameters coming from measuring
$\dot{K}$ and evaluating Eqs.~\ref{newnucrit}-\ref{newvisccrit1} are
also shown in Fig.~\ref{fig:ringsp_cs30}, where $\nu_{{\rm ave}}$ is a
mass-weighted average between $0.5\leq r \leq 1.5$, and $\nusig$ is
the value for the annulus with maximum surface density (same notations
for $\alpha$). For a given flow, both sets of locally-estimated values
of kinematic viscosity and Shakura-Sunyaev $\alpha$ are roughly
similar to the globally-obtained curve-fits, though always having larger
magnitudes.  Comparing the results of the two methods, the values at
the peak surface density of the local calculations are much closer to
those of the global estimates, which is not surprising given that the
former emphasised the maximum-density regions in fitting.  Results
from higher resolution rings were similar but showed a slight
resolution dependence in the effective viscosity values obtained
(discussed further below); a comparison of parameters is given in
Table~\ref{tab:visc_param}.

\begin{table}
  \begin{center}
    \caption{\label{tab:visc_param} Comparison of viscosity parameter
      values at multiple resolutions, from both local (entropic) and
      global (kinematic) methods for the same disc ($\mu_1=30$) shown
      in Fig.~\ref{fig:ringsp_cs30} (at $t=2000$).}
    \begin{tabular}{lllll}
      \hline
      \mc{1}{l}{parameter} &\mc{1}{l}{method} & \mc{1}{c}{$N=5\times10^4$}& \mc{1}{c}{$1\times10^5$}& \mc{1}{c}{$2\times10^5$} \\
      \hline
$h_{\rm ave}$               & -  &  0.058    &  0.045   &  0.036     \\
$\nu_{\rm ave}/(10^{-6})$   & entr.   &  6.0      &  4.6     &  3.5     \\
$\nu_{\rm\Sigma,m}/(10^{-6})$& entr.    &  4.9     &  3.7    &  2.7     \\
$\nu/(10^{-6})$            & kin.    & 3.5       & 2.6      & 2.1      \\
$\alpha_{\rm ave}$         & entr.     &  0.056    &  0.042   &  0.032     \\
$\alpha_{\rm\Sigma,m}$      & entr.    &  0.046    &  0.035   &  0.026     \\
$\alpha$                  & kin.    &  0.03     &  0.02    &  0.02     \\
      \hline
    \end{tabular}
  \end{center}
\end{table}

\begin{figure}
  \includegraphics[width=2.7in]{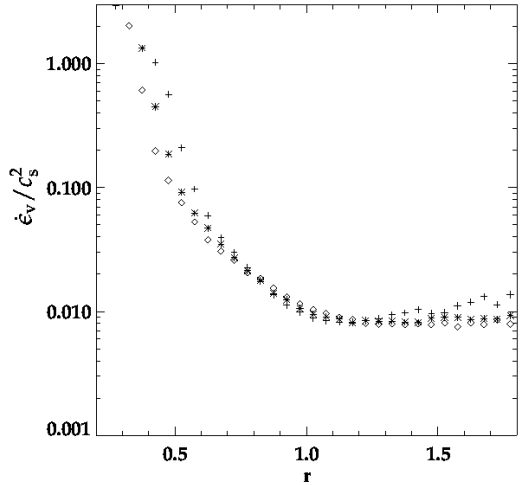}
  \caption{\label{fig:ringsp_cs30N200} The (scaled) specific heating,
    $\dot{\epsilon}_{{\rm v}}/c_{{\rm s}}^2$, at $t=1000$ (`$+$'),
    2000 (`$*$') and 3000 (`$\diamond$') from the same simulation
    shown in Fig.~\ref{fig:ringsp_cs30}. The heating is mostly constant
    for the `outflowing' material at $r>1.0$ but is much higher for
    inner radii.}
\end{figure}

The profiles of the local $\nu$ and $\alpha$ values (lower two panels
of Fig.~\ref{fig:ringsp_cs30}) are roughly constant with time (and
also for the various resolutions tested), with minima near the
location of maximal $\Sigma_{\rm R}$.  This can be understood by
examining the corresponding heating rate profiles in
Fig.~\ref{fig:ringsp_cs30N200}. In the expanding layers at $r>1$,
$\dot{\epsilon}_{{\rm v}}$ is nearly constant, and therefore, by
Eq.~\ref{newnucrit}, $\nu \propto r^3$.  The inner layers are
compressed as they spread and have a much higher heating rate.  In the
regions containing the most mass ($0.7\leq r\leq1.3$), the values of
viscosity and the Shakura-Sunyaev parameter are fairly constant, in
accord with the overall model assumptions.

The large viscosity values at the inner and outer disc boundaries
(Fig.~\ref{fig:ringsp_cs30}) correspond to very low density regions
which are poorly resolved.  A consequence of the smoothing length
condition in Eq.~\ref{sph_hsml}, used to stabilise the interpolations,
is that particle kernels are artificially large at the edges of the
simulation, and therefore the density is artificially low.  Hence,
Gadget-2 and other SPH formulations with similarly defined smoothing
lengths necessarily require careful boundary conditions. As a function
of time, the values of both $\aave$ and $\asig$, increase slightly
(along with $\have$).

\begin{figure*}
    \begin{tabular}{cc}      
      \includegraphics[width=2.2in]{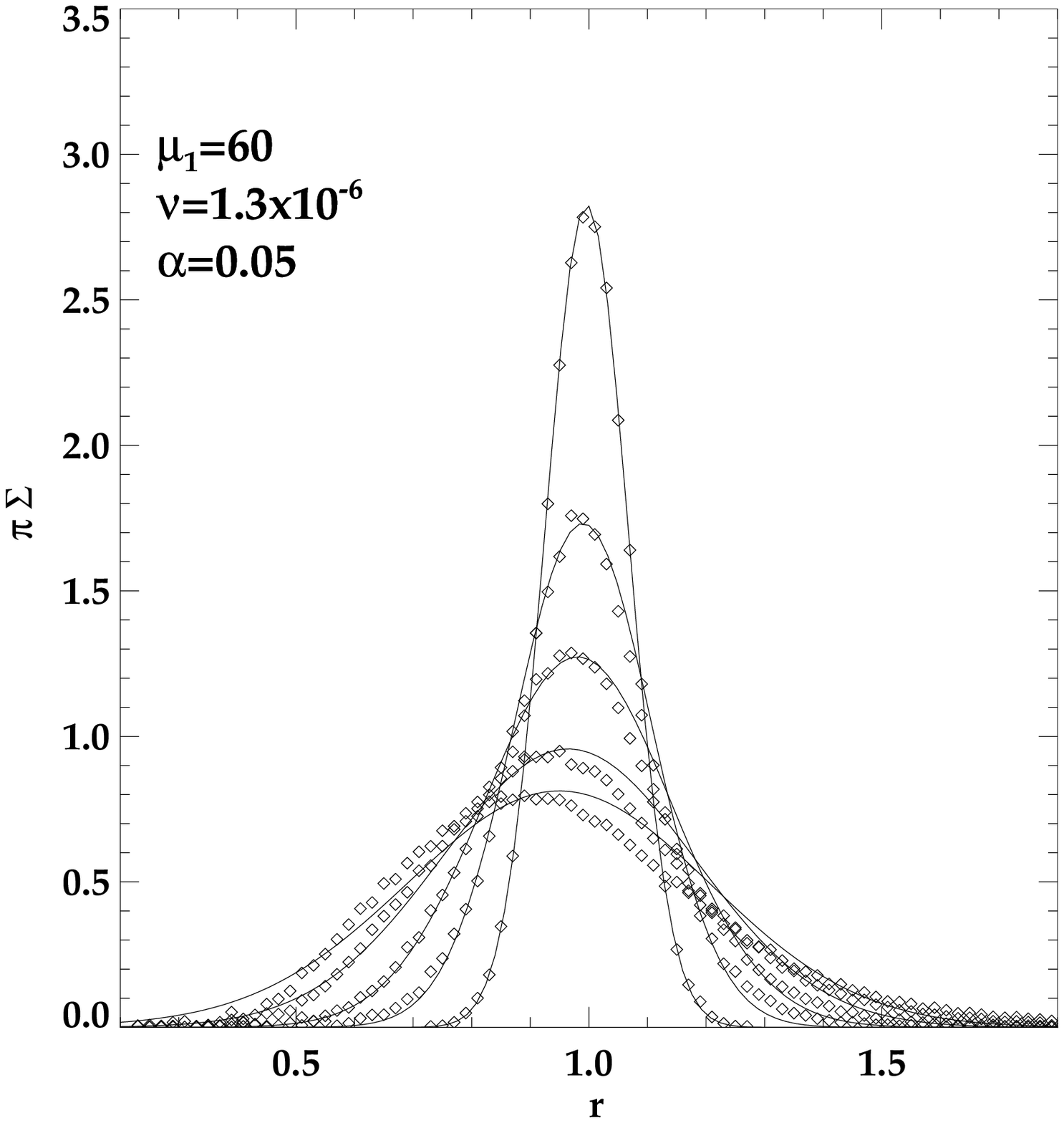}&           
      \includegraphics[width=2.2in]{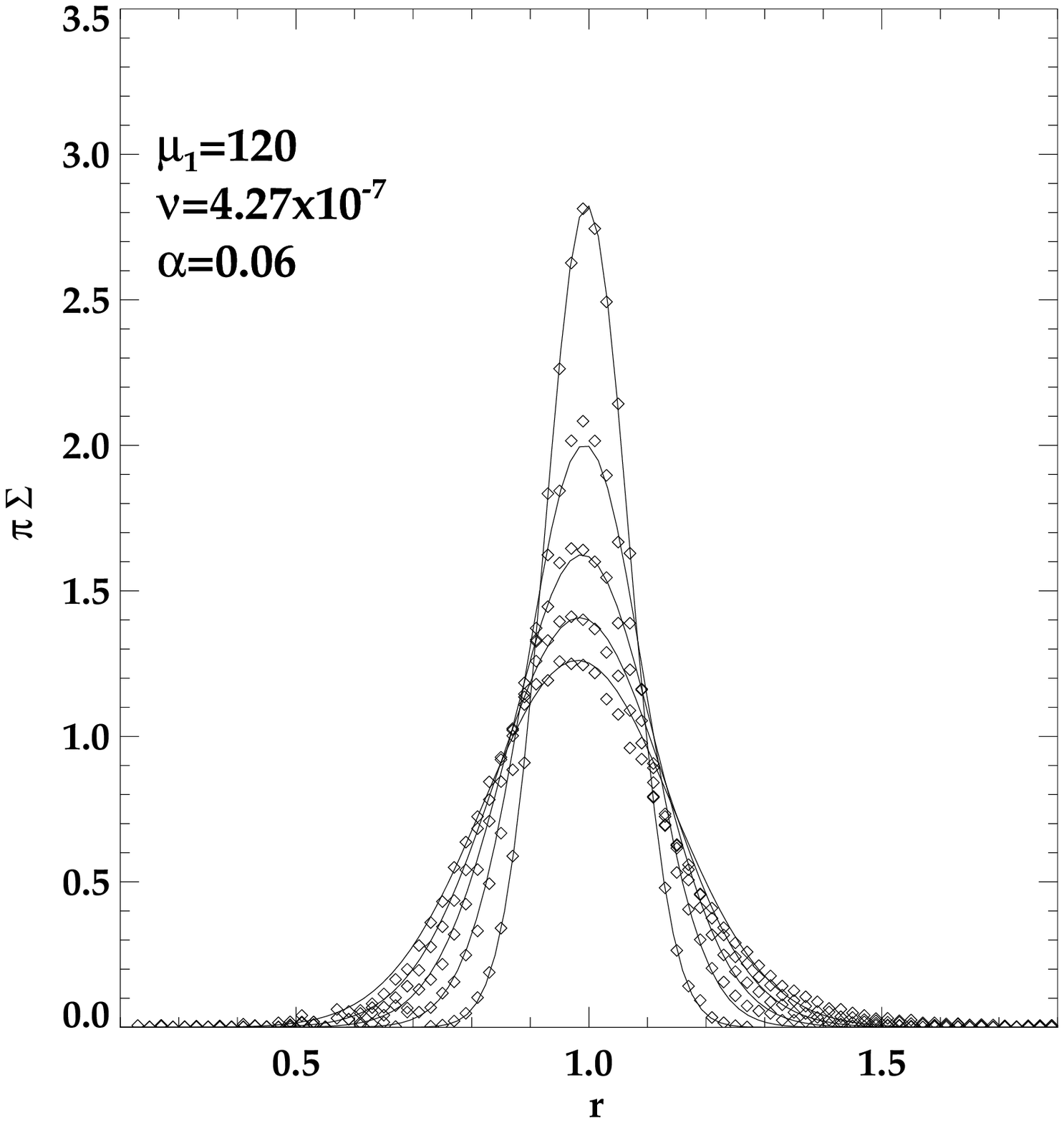}\\
      \includegraphics[width=2.2in]{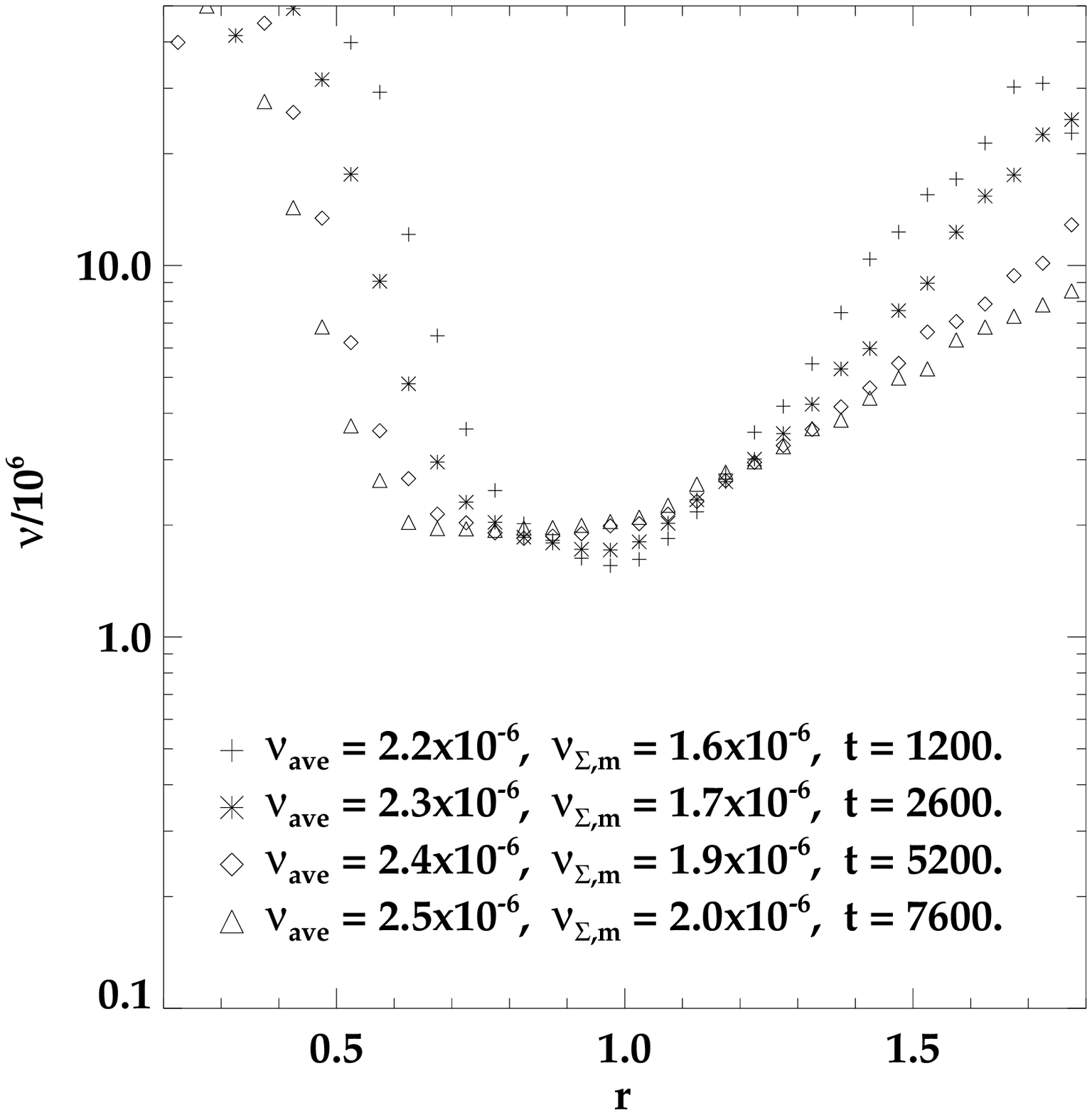}&           
      \includegraphics[width=2.2in]{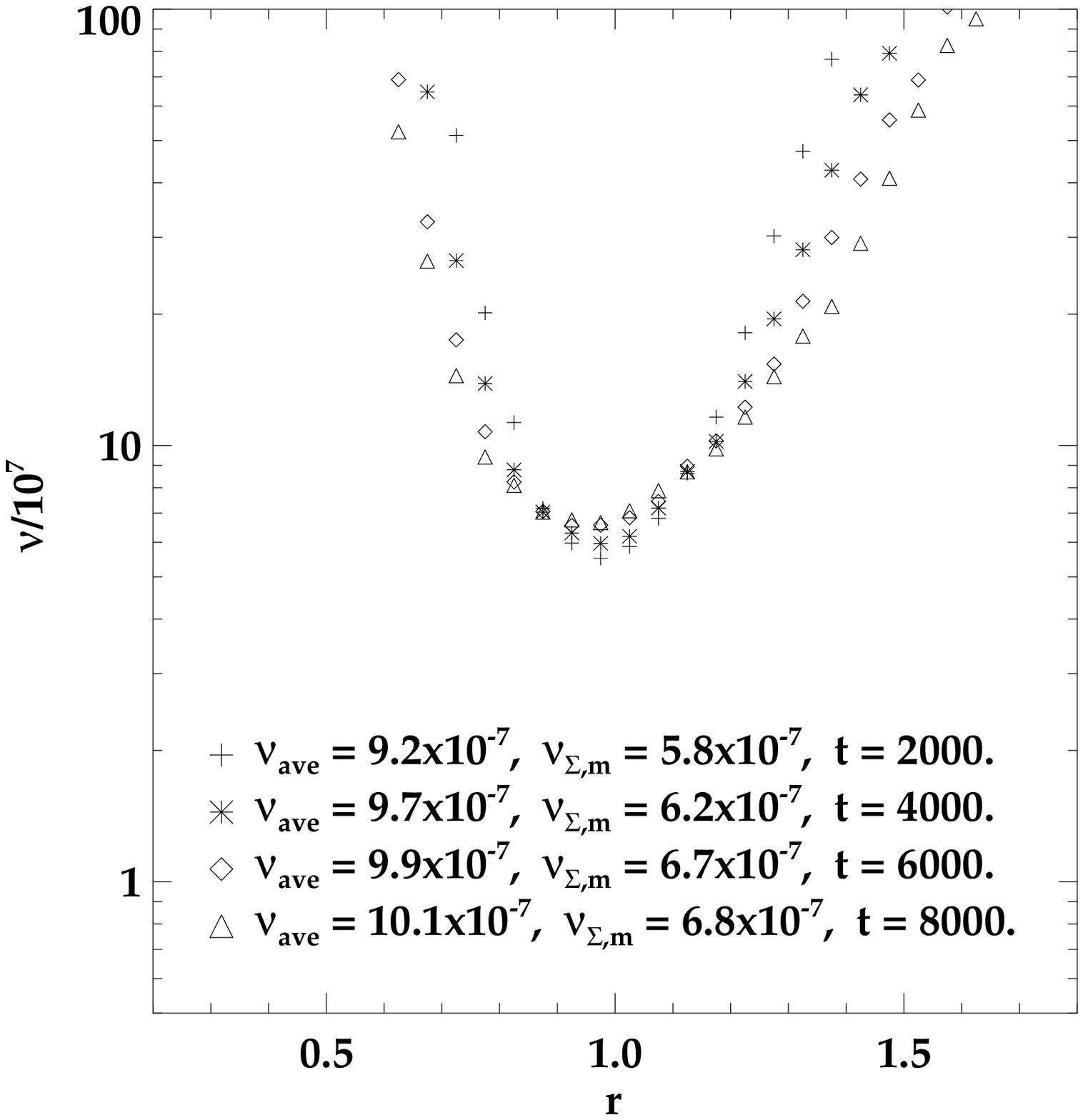}\\
      \includegraphics[width=2.2in]{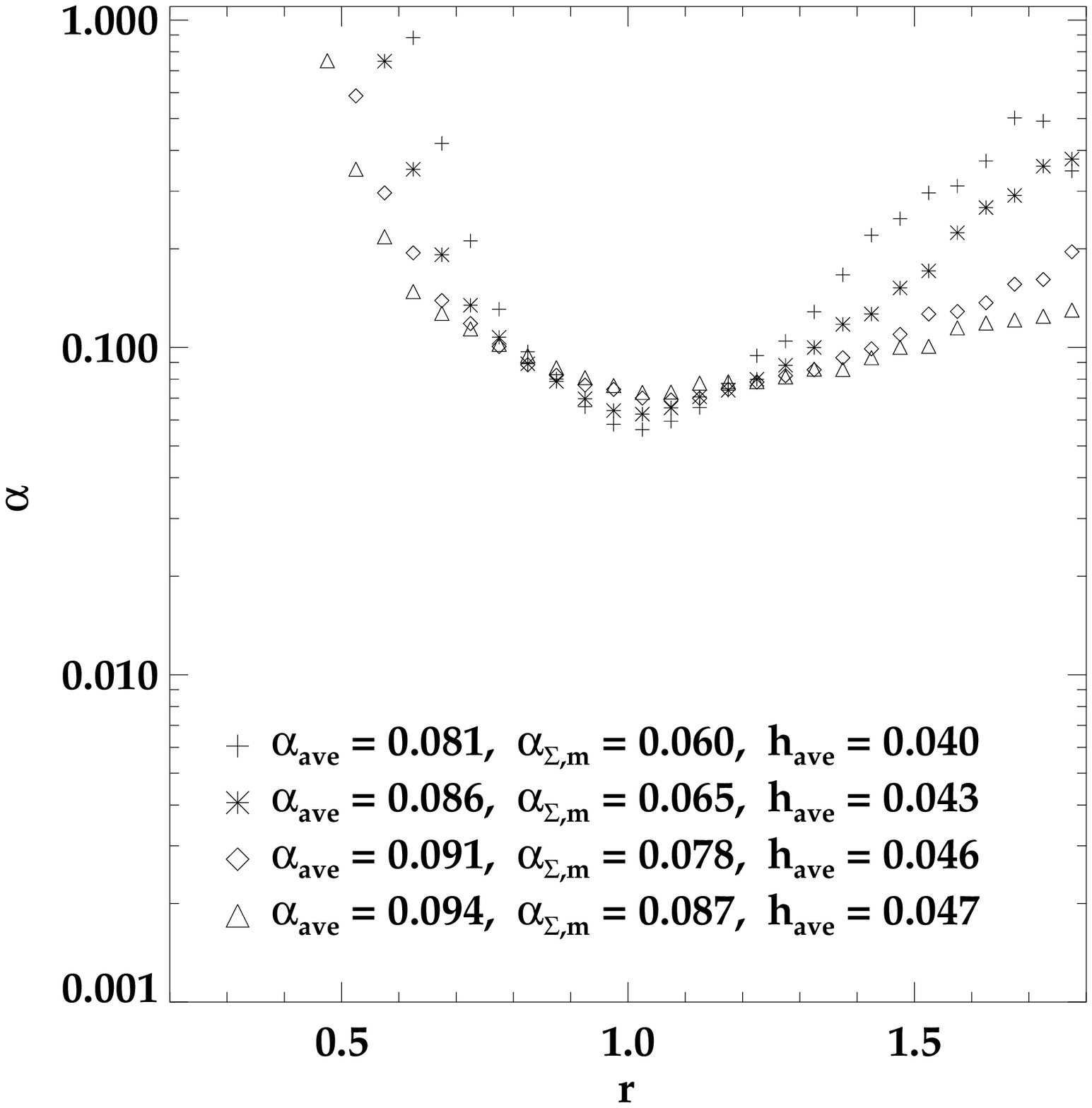}&           
      \includegraphics[width=2.2in]{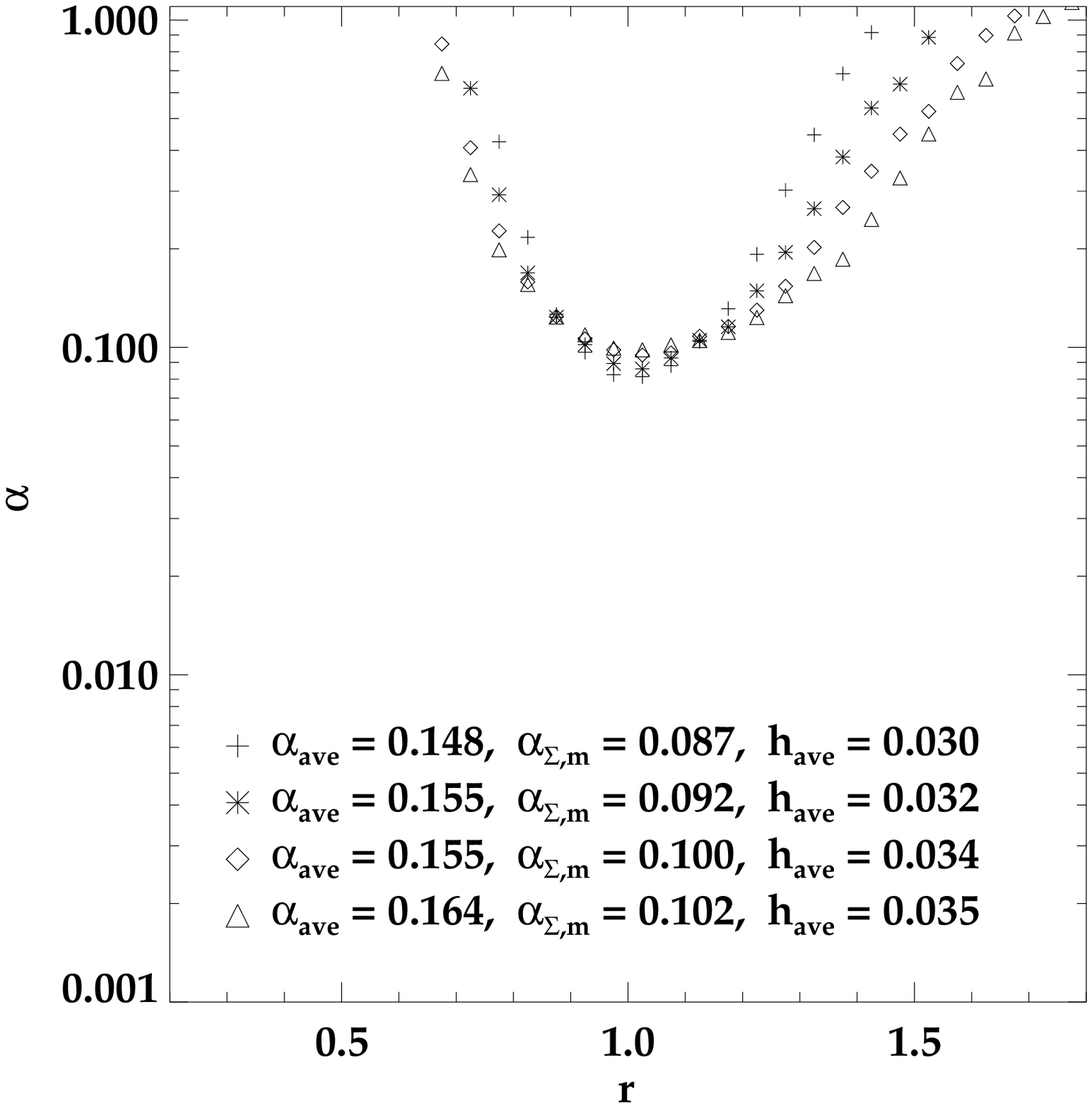}            
  \end{tabular}
  \caption{\label{fig:ringsp_cs60} Ring-spreading test with $N$=50k,
    shown at times given in the middle panel (and at $t=0$, top).
    Left, $\mu_1=60$; right, $\mu_1=120$
    (cf. Fig.~\ref{fig:ringsp_cs30} for notations).}
\end{figure*}

Fig.~\ref{fig:ringsp_cs60} shows results from investigating other
values of $\mu_1$ ($=60$ and 120), again using $5\times10^4$
particles.  For both the global/kinematic and local/entropic measures,
$\alpha$ increases with increasing $\mu_1$, whereas $\nu$ decreases
(with the local/entropic values being systematically higher than the
global/kinematic ones in each case).  The resolution trends are the
same as those for $\mu_1=30$.  It then appears that, while the value
of the artificial viscosity parameter has remained constant ($\av=0.8$
for all of these simulations), the kinematic viscosity and
disc-$\alpha$ depend on both the sound speed and the resolution.  To
explain the former dependence, one can note that $c_{{\rm s}}$ appears
directly in the expression for $\Pi_{ab}$ in Eq.~\ref{visc_sph}.  The
latter dependence results from the fact that, inherently in SPH, the
smoothing length of a fluid element determines the length scale of its
interactions.  With decreasing particle number, the cross-sectional
spread of an element increases ($\have^2\propto N^{2/3}$), resulting
in greater shear in the flows.  Using only the bulk term in $\Pi_{ab}$
for studying non-axisymmetric discs, \citep{1996MNRAS.279..402M}
estimated linear dependence for both sound speed and smoothing length
(which was assumed to be constant), i.e., $\nu\propto c_{{\rm s}}h$;
recently similar analysis has been performed for the case of
artificial tensor-viscosity in studies of warped discs by
\citep{2010MNRAS.405.1212L}.

This dependency of effective viscosity on features which vary across
the disc shows the importance of local measurement. Both the spatial
and temporal variations in surface density away from the exact
analytic solution have been noted above.  In using the local estimates
to determine parameters with Eqs.~\ref{newnucrit}-\ref{newvisccrit1},
one has the dual advantage of avoiding any arbitrariness which arises
in global curve-fitting and of characterising radial dependences
quantitatively.  This approach allows a more complete description of
the behaviour of the SPH artificial viscosity in the ring.

For example, in the very thin $\mu_1=120$ rings the effective
viscosity has a much smaller constant-value region than in rings with
higher $c_{{\rm s}}$.  Profiles of $\alpha$ steeply increase to either
side of $r\approx1$, also making analytic curve-fitting less exact.
Apart from differences due to compression and expansion of annuli, the
curves reflect the influence of boundary conditions and the
effectively poorer resolution within these flattened discs
($H\propto\mu^{-1}$), even though the volume is smaller for the same
$N$: since the ratio of surface area to volume is necessarily larger
as well, the kernels of more SPH particles encompass empty space
outside the disc, artificially increasing $h$ (and similarly
decreasing $\rho$). This would suggest the practical necessity for
increased $N$ in thinner, high-$\mu$ discs to avoid numerical effects
in the viscous evolution and artificially high accretion rates.  In
some sense, this is a statement of the intuitive notion that a
resolution criterion such as $H/h\geq$ (a few) is required for
accurately characterizing discs.

\begin{figure*}
    \begin{tabular}{cc}
      \includegraphics[width=2.2in]{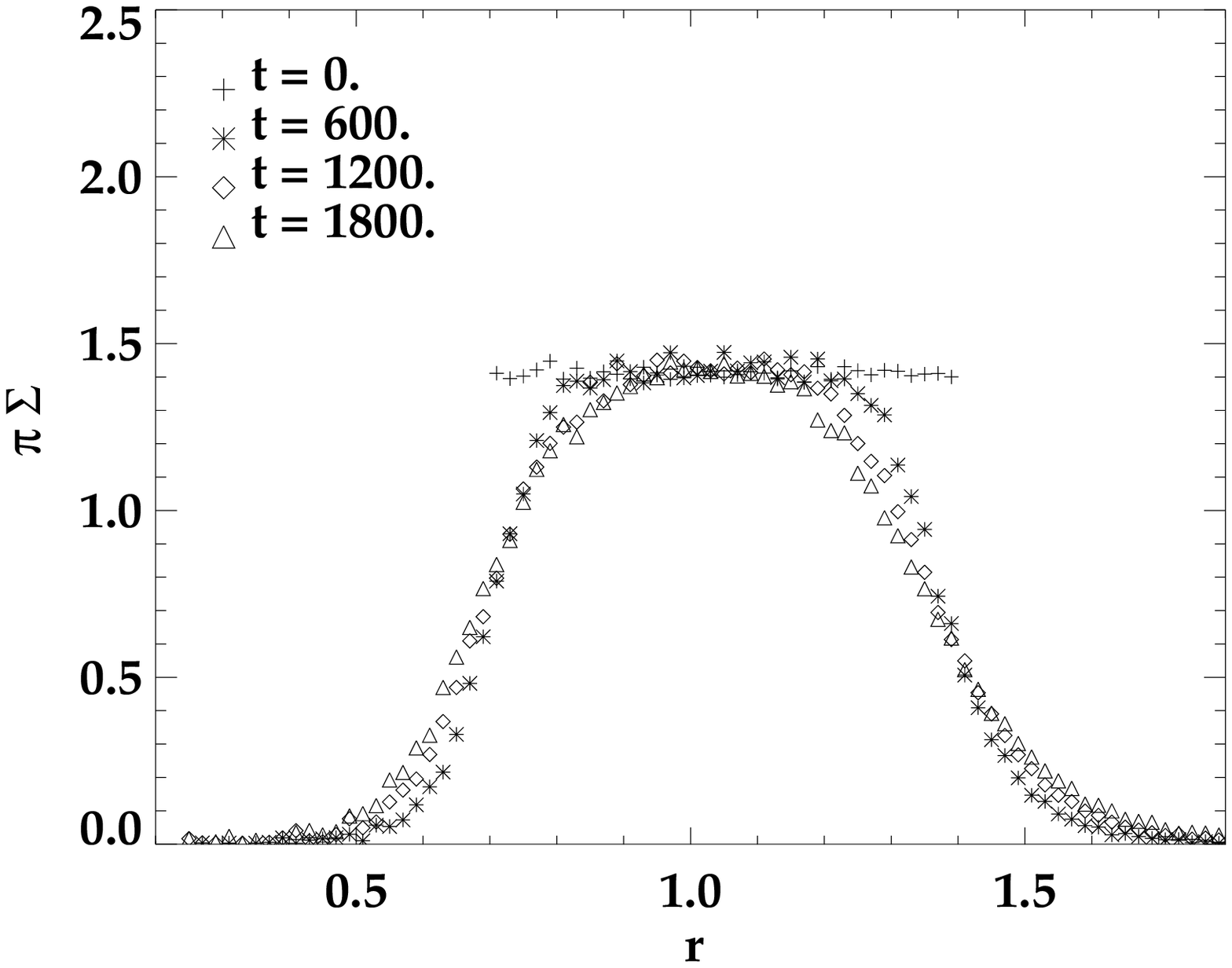}&           
      \includegraphics[width=2.2in]{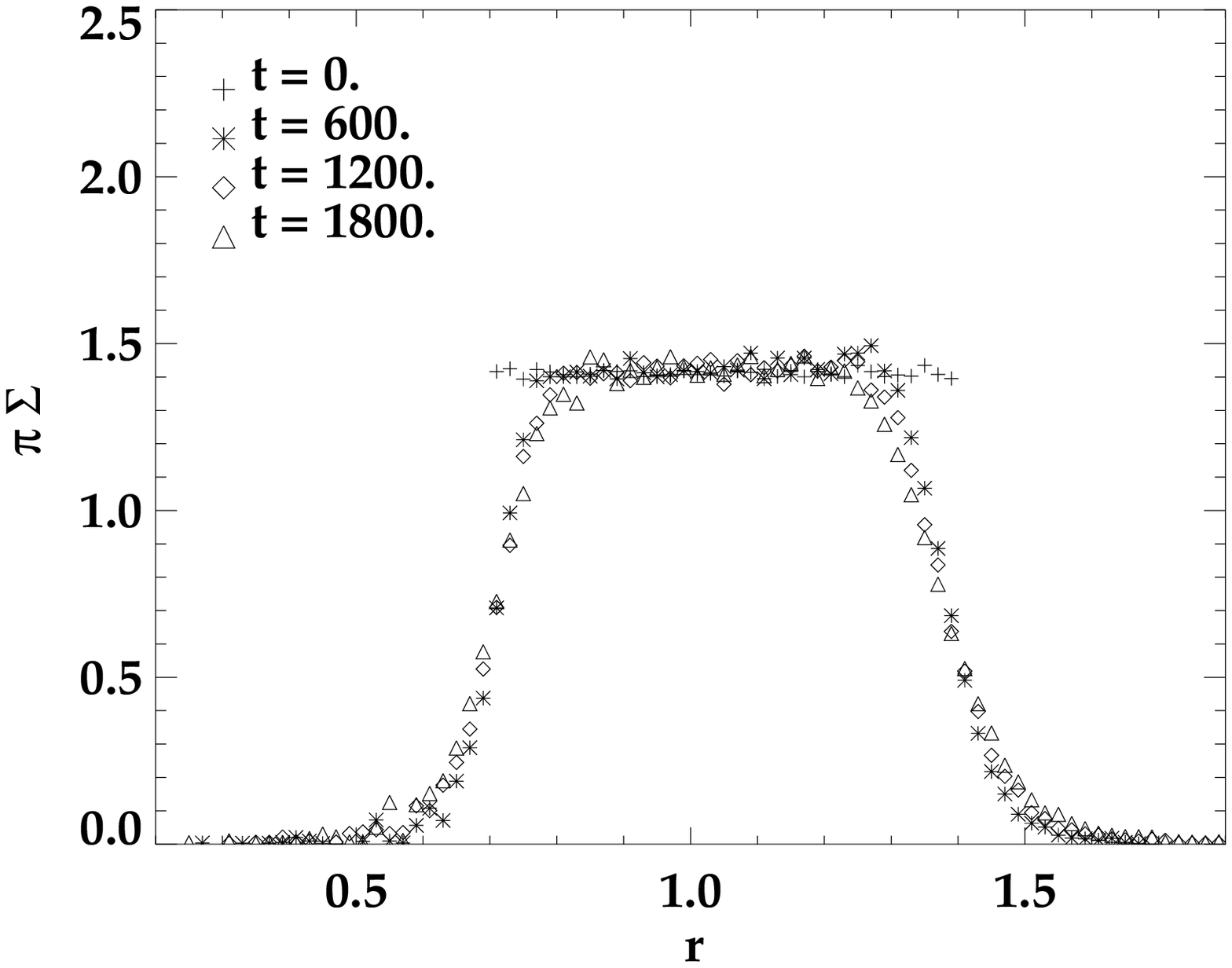}\\
      \includegraphics[width=2.2in]{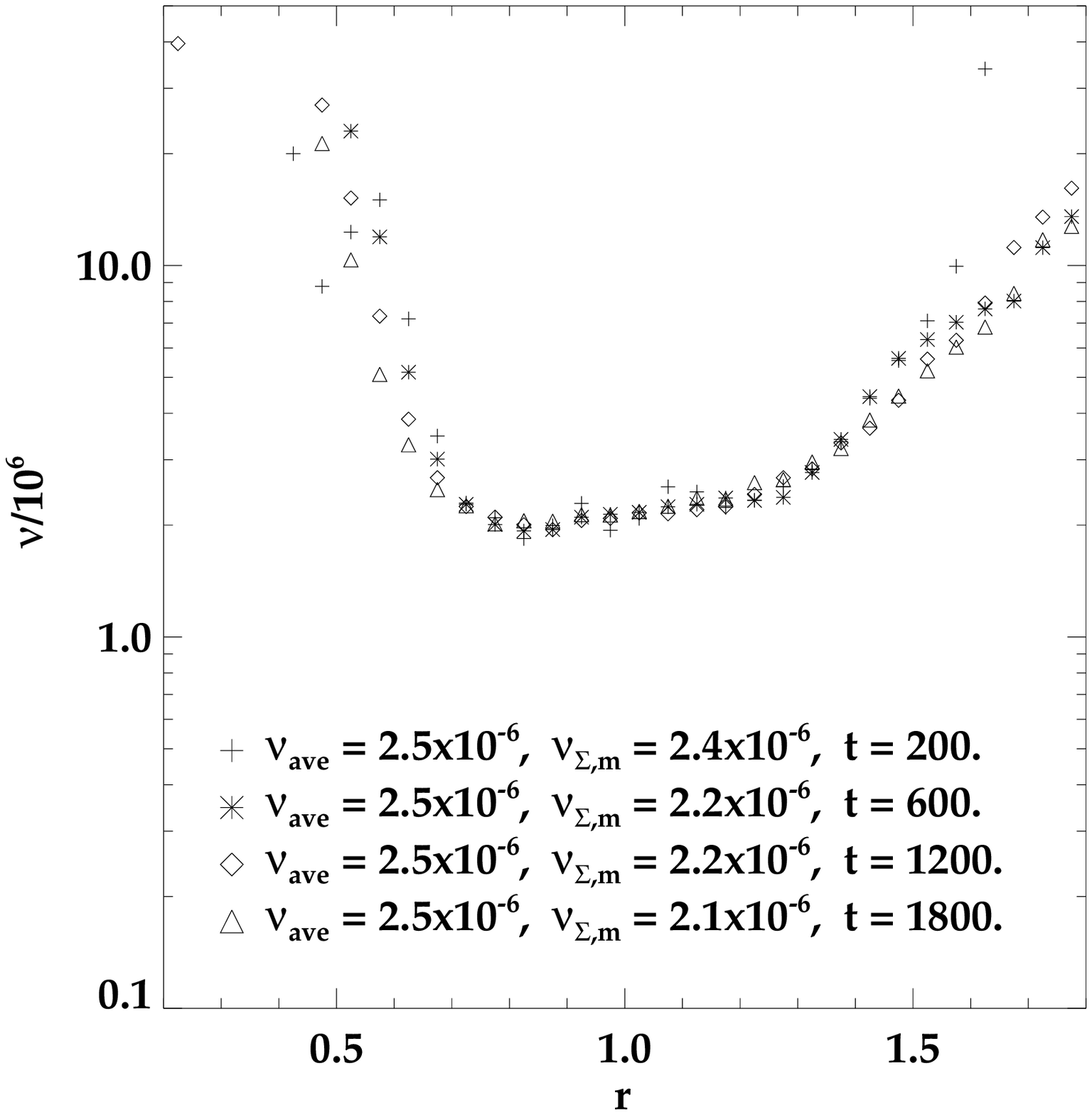}&           
      \includegraphics[width=2.2in]{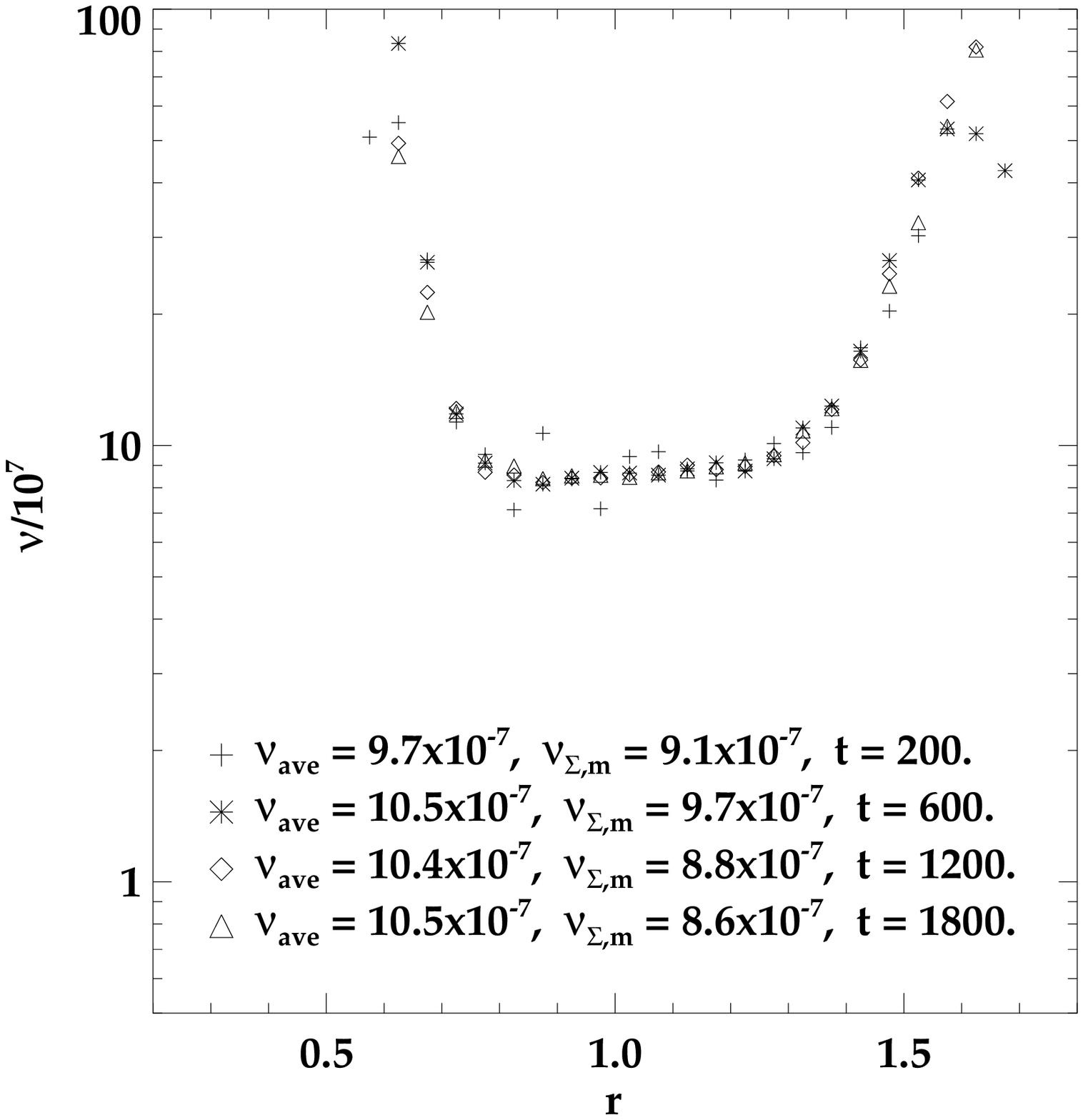}\\
      \includegraphics[width=2.2in]{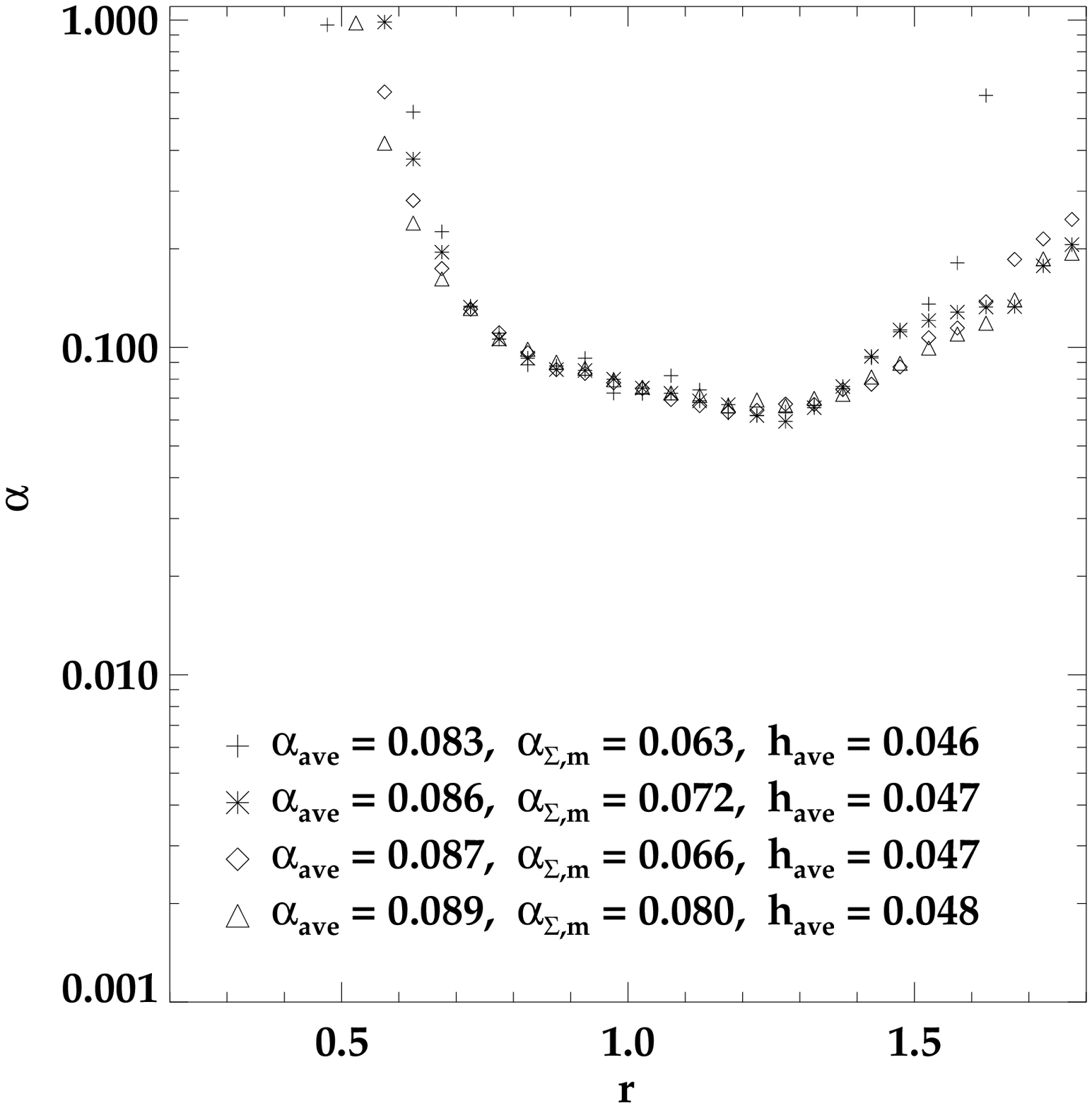}&           
      \includegraphics[width=2.2in]{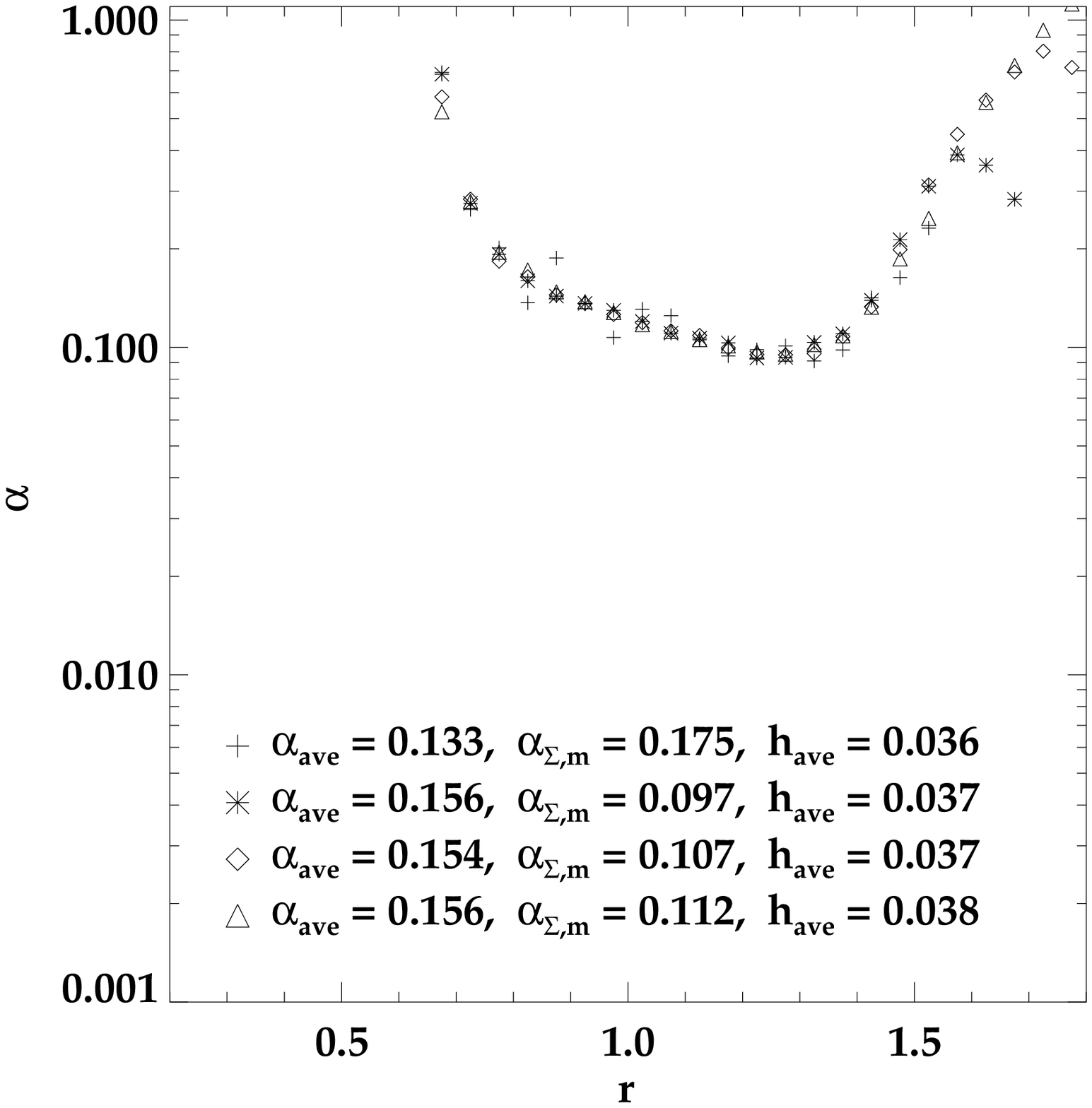}            
  \end{tabular}
  \caption{\label{fig:discsp_cs60} Flat-profile discs with constant
    $\Sigma_{{\rm F}}(x,t=0)$ (as described in
    \S\ref{sec:thindiscstr}) and $N=5\times10^4$. Left, $\mu_1=60$;
    right, $\mu_1=120$.}
\end{figure*}

As a comparison, we now investigate rotating flows with flatter
surface density profiles, for which the `empty kernel' boundary
behaviour should have less effect on the evolving system.  We
therefore use local viscosity measures to analyse the systems with the
$\Sigma_{{\rm F}}$ defined in \S\ref{sec:thindiscstr}, which are
initially more `disc-like' in the sense of having a greater radial
extent of mass.  Time-evolution profiles of surface density are shown
in Fig.~\ref{fig:discsp_cs60} for $\mu_1=60$ and $\mu_1=120$.
Indeed, for these discs the values of $\nu$ and $\alpha$ are fairly
constant functions of radius, and, for a given $\mu_1$, the measured
viscosity values agree well with those from the ring tests.

\begin{figure*}
  \begin{tabular}{cc}
    \includegraphics[width=3in]{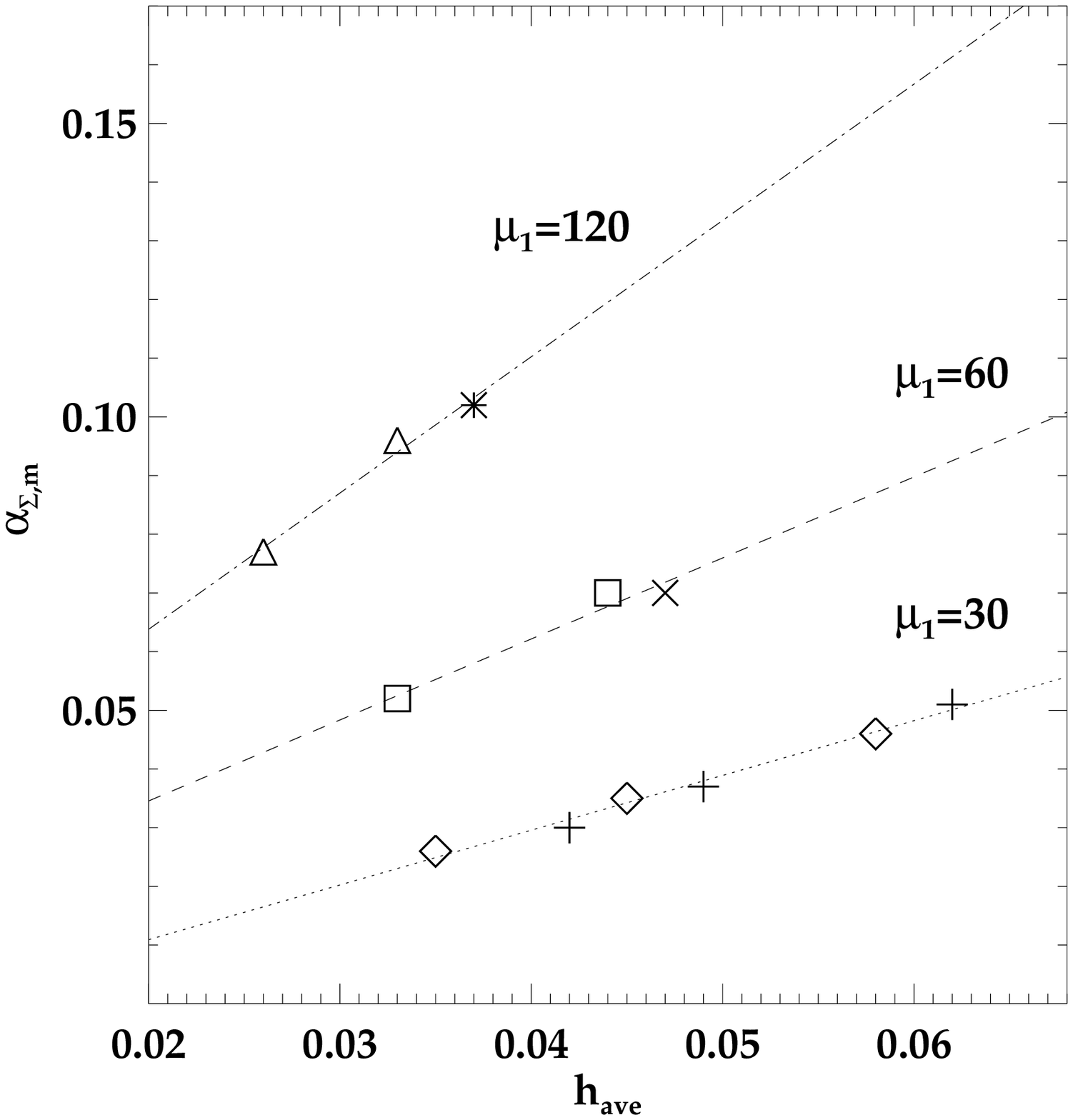}&           
    \includegraphics[width=3in]{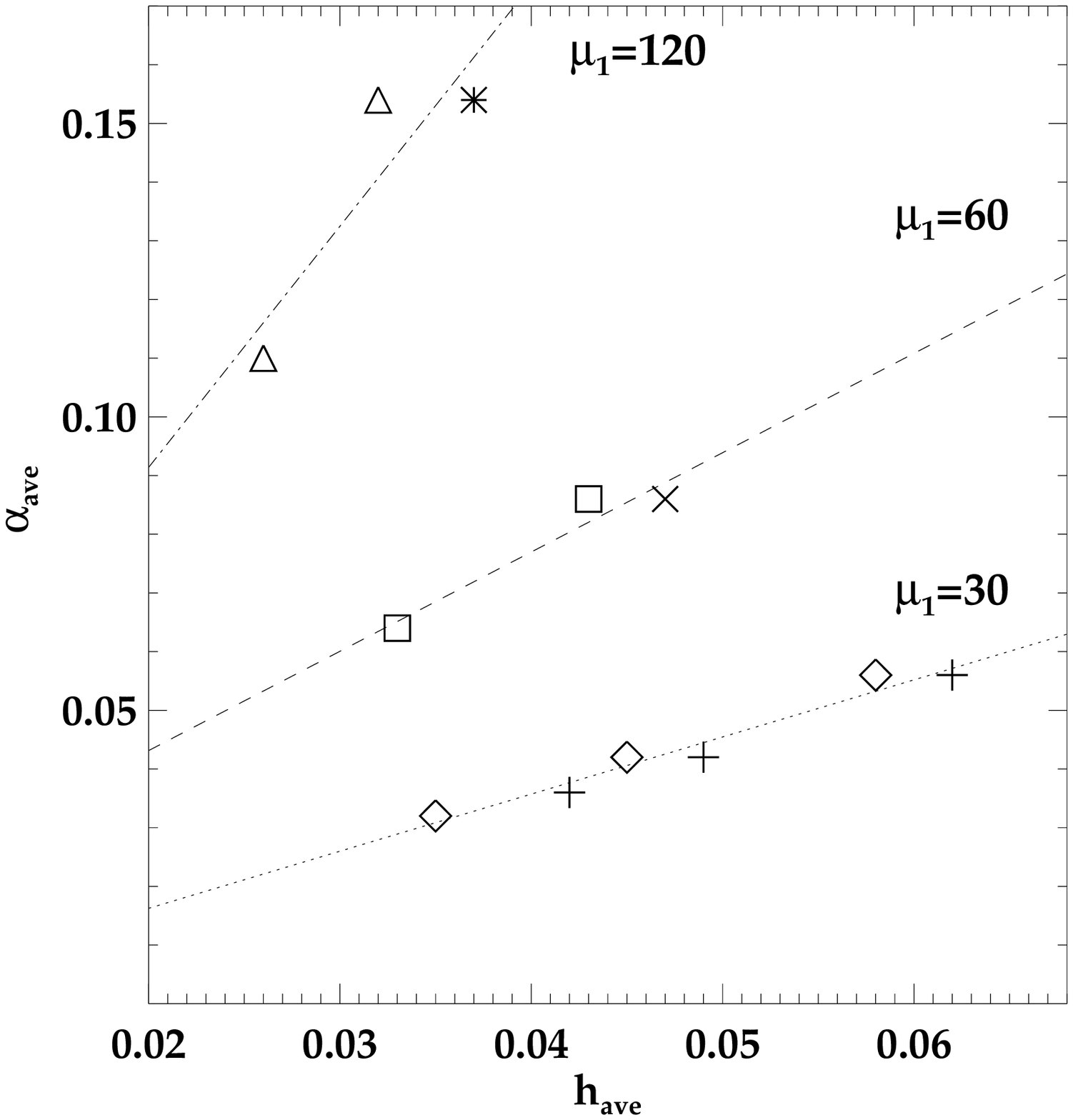}\\
  \end{tabular}
  \caption{\label{fig:phasesp} Phase space of $\asig$ (left) and
    $\aave$ (right) vs $\have$ from all of the rotating flow
    simulations in this study using $\Pi'_{ab}$ and $\av=0.8$. Ring
    and flat-disc simulations ($\Sigma_{{\rm R}}$ and $\Sigma_{{\rm
        F}}$) are represented, respectively, by `$\diamond$' and `+'
    for $\mu_1=30$; by `$\square$' and `$\times$' for $\mu_1=60$; and by
    `$\triangle$' and `$\ast$' for $\mu_1=120$.  The points appear
    aligned by azimuthal mach number, and best fit lines for
    $\mu_1=30,\,60,\,120$ are included as visual guides. }
\end{figure*}

In order to quantify the behaviour of the Balsara-corrected artificial
viscosity prescription, $\Pi'_{ab}$ with a given $\av$, for rotating
flows with different resolutions and sound speeds, the resulting
Shakura-Sunyaev parameter values for all rings and discs from this
study were plotted in $\have$-$\aave$ and $\have$-$\asig$ planes
(Fig.~\ref{fig:phasesp}).  Note that the $\alpha$ values appear to
vary quite linearly with smoothing length, with $\mu_1$ determining
the gradient; the values of $\asig$ appear to be independent of the
specific profile of surface density, whereas disc values have
noticeably lower values of $\aave$ than the rings.  This (small)
qualitative difference is to be expected, as boundaries affect the
former less.  While the values in both plots appear to be
quasi-convergent towards $\alpha=0$ as $\have$ tends to zero, it is
not possible to be precise about this due to the approximate nature of
the plot.

\subsection{Further examples and altering artificial viscosity}

\subsubsection{Evaluating the artificial viscosity in Balsara-corrected flows}

The simulations presented above have all utilised the same value of
$\av=0.8$ and the Balsara-corrected form for the artificial viscosity,
$\Pi'_{ab}$.  In this section we investigate the effects of varying
$\alpha_{\rm v}$, using only the local entropy production method to
measure the viscous properties of the flows.  It should be noted that
for these predominantly smooth and non-shocking flows, the linear term
of the artificial viscosity dominates entropy production.  Neglecting
the quadratic term ($\bv=0$) produces very little change in the disc
flow.  Therefore, we only further consider here the effects of varying
$\av$ and of using or not using the Balsara correction.

With $\Pi'_{ab}$, even significantly different values of $\av$ result
in only minor changes of the disc parameters, as was to be expected.
Originally, the ring with $\mu_1=30$ and $\av=0.8$ (seen in
Fig.~\ref{fig:ringsp_cs30}) had ($\aave,\,\asig$) values of
approximately (0.056, 0.046); increasing to large values of $\av=1.6$
and 2.4 resulted in (0.060, 0.049) and (0.066, 0.057), respectively;
only at extremely high values were significant changes seen,
e.g. $\av=8.0$ yielded (0.109, 0.091).  Thus, in these purely
rotational flows, the shear-correction does limit the artificial
viscosity term (although we stress again that it remains significantly
non-zero), and the values shown in Fig.~\ref{fig:phasesp} are roughly
independent of $\av$ within the range studied.  From fitting various
disc parameters for results from the preceding simulations (across a
range of radii, $0.5<r<1.5$), it emerges that the effective viscosity
is well represented by the following (simple) expression:
\begin{equation}\label{alphalphrelbal}
  \alpha(r) \approx \frac{1}{16}\left[\frac{h(r)}{2H(r)}\right]\,,
\end{equation}
explicitly noting that $\alpha$ generally varies with position, as
both the smoothing length and disc height are functions of
radius. Therefore, when the Balsara correction is applied, the
interaction length among particles appears to strongly determine the
viscous dissipation in the flow.

\begin{figure}
    \includegraphics[width=3in]{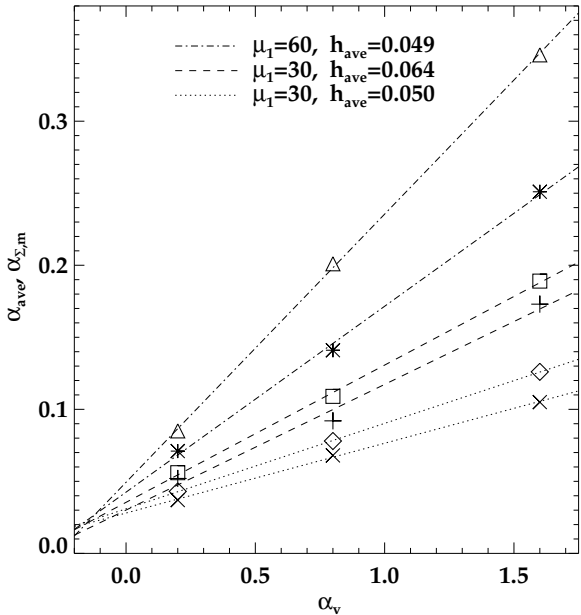}         
  \caption{\label{fig:phasespPi} Phase space of Shakura-Sunyaev alpha
    values vs SPH artificial viscosity, using the non-Balsara
    corrected $\Pi_{ab}$.  For $\av=0.2,\,0.8,\,1.6$, $\aave$ values
    are respectively shown in open symbols ($\triangle$, $\square$ and
    $\diamond$) and $\asig$ values, in skeletal symbols (`$\ast$', `+'
    and `$\times$'). The different values of $\mu_1$ and $\have$ (the
    latter mainly varying with $N$) are shown on the figure, with best
    fit lines of the data points as visual guides.}
\end{figure}

\subsubsection{Evaluating the uncorrected artifical viscosity in flows}

\begin{figure}
    \includegraphics[width=3in]{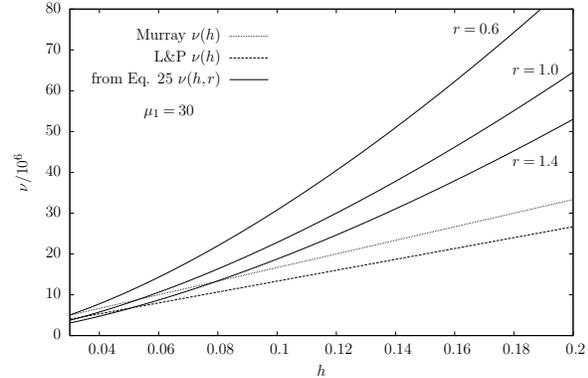}         
 \caption{\label{fig:nu_vs_h_compar} A comparison of analytic
   expressions for the effective disc viscosity, $\nu$.  Previous
   analytic expressions from \citet{1996MNRAS.279..402M} and from
   \citet{2010MNRAS.405.1212L} are shown with dotted and dashed lines,
   respectively; solid lines show Eq.~\ref{alphalphrel} for
   representative radial values.  In this ring simulation with
   $N=5\times10^4$, the $h$ values were mainly between
   $3\times10^{-2}$ and $2\times10^{-1}$, hence the displayed interval
   of the Eq.~\ref{alphalphrel} curves.}
\end{figure}

In switching from $\Pi'_{ab}$ to $\Pi_{ab}$ (no Balsara correction),
the effective viscosity for $\av=0.8$ is nearly doubled.  The clearly
linear dependence of $\asig$ and $\aave$ on $\av$ in this case (and
with $\bv=3\av$) is shown for a series of discs in
Fig.~\ref{fig:phasespPi}.  Best fit lines of the data points are
provided as visual guides, with open and skeletal symbols denoting
$\aave$ and $\asig$, respectively.  The dependences on both $c_{{\rm
    s}}$ and $\have$ (which is mainly determined by $N$) remain
qualitatively similar to those observed in the Balsara corrected
simulations (i.e., increasing either parameter leads to an increase in
$\alpha$).  In the range of these examples, the `empirical'
dependence of the Shakura-Sunyaev $\alpha$ on SPH and flow parameters
can be derived from a fitting procedure similar to that used for
Eq.~\ref{alphalphrelbal} above and given as:
\begin{equation}\label{alphalphrel}
\alpha(r) \approx 0.43\,\av\,\mu(r)^{-1/3}
\left[\frac{h(r)}{2H(r)}\right]^{3/2}\,.
\end{equation}
The functional form offers direct insight into the dependences of the
effective viscosity on disc parameters, and this relation can be
compared with existing approximate, analytic formulae in the
literature.  Fig.~\ref{fig:nu_vs_h_compar} shows a comparison of $\nu$
derived from Eq.~\ref{alphalphrel} with formulae derived by
\citet{1996MNRAS.279..402M} and \citet{2010MNRAS.405.1212L}.  Even
though the latter two relations were applied for different SPH schemes
(as well as containing some dependence on both differing kernels and
dimensionality), the figure shows that there are regions of good
agreement with the empirical values, namely for small smoothing
lengths.  (We note that the curves derived in the present study have
been restricted to the range of $h$ values observed in simulations.)
However, as $h$ increases, the differences from the analytic relations
increase significantly.

Formulae such as Eq.~\ref{alphalphrel} can be useful in efficiently
establishing the setup of a simulation (e.g., choosing the number of
particles to produce a required smoothing length, as a sort of
resolution requirement). For the non-Balsara flows, extrapolation far
beyond the examined parameter space with any approximated function is
most likely unsuitable. However, a function such as this coming from
tests with a fairly wide range of values for the independent
quantities may provide an adequate level of accuracy for
interpolation.

\section{Summary and discussion}
\label{sec:discuss}

An entropy-based local method has been described to measure the
effective kinematic viscosity in the flow due to SPH artificial
viscosity, which can be simply related to the Shakura-Sunyaev $\alpha$
for appropriate thin discs.  This approach has several advantages over
that of estimating a global $\nu$ from an isothermal ring-spreading
test.  First, it calculates the local viscosity of an annulus from the
(viscous) energy or entropy expression, which contains only the term
dependent on $\Pi_{ab}$, and therefore only viscous effects are
measured.  This procedure then yields $\nu$ and $\alpha$ as functions
of radius, so that variations in the effective values across a disc
can be determined; global values may be obtained easily via averaging,
if desired.  In addition, results are generated immediately, without
the necessity of finding fits for analytic curves as in
ring-spreading, and also thereby eliminating arbitrariness from the
estimates. Moreover, the viscosity values are calculated from
quantities which are already calculated in the SPH algorithm.

Finally, whereas traditional ring-spreading tests are restricted to
use with an isothermal equation of state (or, in practice, even more
restricted to cold, 2D systems) because it is only then that one has
the analytic solution for comparison, this local method can be used
with any equation of state.  The effective viscosity may be obtained
for rotating flows in general as a function of viscous dissipation,
such as from Eq.~\ref{general_visc}or \ref{nu_szmil} (both derived
using the very general Eq.~\ref{visc_en_entr}). The exact form would
simply differ from that of Eq.~\ref{newnucrit}, in which a Keplerian
profile ($\Omega = \Omega_{{\rm K}}$) has been assumed.  Further work
will also investigate an analogous approach applicable to estimating
the effective local shear viscosity for systems in linear motion.

We have compared two existing methods (standard ring-spreading and
analytic approximations) with the new, local entropy method in
studying the behaviour of the SPH artificial viscosity in rotating
flows. In cases where the particular assumptions of the former methods
were applicable, average results of all approaches were shown
generally to be in fair agreement, with far greater detail coming from
the local entropy method. Further examples changing the disc structure
and the SPH artificial viscosity values were given.  Examples of
explicit expressions which approximate the dependence of the numerical
$\alpha$ on various flow and SPH parameters were given for both the
Balsara and non-Balsara corrected forms of artificial
viscosity. Procedures such as these may provide useful guides when
setting up simulations.  However, further work must be done to test
the generalizability of these specific forms, i.e.
Eqs.~\ref{alphalphrelbal}-\ref{alphalphrel}, across wider parameter
space and to other SPH codes. The means for estimating $\nu$ and
$\alpha$ in each case, however, applies generally with the local
method.

We note that the relations between SPH variables and the effective
viscosity in disc flows described here can serve as useful examples of
the type of analysis possible with the local entropy measures.  The
ability for hydrodynamic schemes to represent flows with particular
physical viscosity characterizations, such as the widely-used
Shakura-Sunyaev discs, may be investigated more directly and in more
detail with the local method.  Moreover, this analysis may be useful
in the further development and evaluation of physically-motivated
viscosity terms in SPH, to which existing methods cannot typically be
applied. For example, it was shown here that the $\alpha$ in the case
of Balsara corrected artificial viscosity remained finite and linearly
dependent on smoothing length. Studies continue to develop schemes for
minimizing unwanted effects of artificial viscosity in rotating flows,
such as the recent work by \citet{2010MNRAS.408..669C}, and we suggest
that the simple approach outlined here for measuring the local viscous
effects directly would be extremely useful in quantitatively assessing
these developments. (Indeed, we will return to discussion of this in a
following paper.)

Finally, several studies have described and investigated numerical
implementations of physical viscosity prescriptions derived directly
from the analytic Navier-Stokes equations (e.g.,
\citet{1994ApJ...431..754F,2006A&A...453.1027L}), with separate bulk
and shear viscosity terms (unlike the standard SPH artificial
viscosity, which does not differentiate the two); this is what one
must do in order to incorporate real viscosity behaviour into SPH
calculations. However, some numerical considerations remain for such
physical viscosity prescriptions, in particular regarding spatial
derivatives and dependence on particle disorder (see, e.g.,
\citet{2009NewAR..53...78R}, as well as Appendix \ref{appendixa} of
the present paper). While one could think of using the SPH artificial
viscosity as a model for a real, physical one (see, e.g., the
discussion in Sec. 3.2.3. of \citet{2010MNRAS.405.1212L}), this
provides only a rough approximation, and it is best to think of the
present work in terms of establishing acceptable limits for the setup
parameters so that the artificial viscosity does not interfere with
the effects of the real one.

\section{Acknowledgements}

The authors thank Scott Kay, Shazrene Mohamed, S\'{e}bastien Peirani,
Philipp Podsiadlowski and Stephan Rosswog for useful discussions
concerning Gadget-2 and SPH. Many of the computations reported here
were performed with the computing cluster at the Centre de Calcul
Scientifique at the Universit\'{e} de Sherbrooke, and we thank Lorne
Nelson for enabling us to use this. We also gratefully acknowledge
support from CompStar, a Research Networking Programme of the European
Science Foundation.



\appendix
\section[A]{Initial particle distribution}
\label{appendixa}
As any SPH simulation evolves, the particles asymptotically tend
toward an equilibrium of mutual separation distances (i.e., no
artificial clumps or voids) within the local profile
\citep{2006MNRAS.365..199M}.  The presence of either
lattice-regularity or purely random (over)density fluctuations in the
initial conditions impedes this process, resulting instead in purely
numerical features, the effects of which may continue to propagate in
the simulated flows, e.g.,
\citet{2002ApJ...569..501I,2006MNRAS.365..199M}.  For example,
locating SPH particles at regular gridpoints often leads to spurious
spiral modes in rotating systems (though `real' spirals may form due
to susceptibility conditions in certain cases, such as those given by
\citet{1964ApJ...139.1217T,2003A&A...399..395S}).  In particle setup
methods using random placements of equal-mass particles, some of these
numerical side effects are limited, but fluctuations may still occur
in the flow which are of an entirely artificial nature, particularly
due to localised overdensities; typical solutions of these
difficulties have been to enforce a minimum initial separation between
randomly placed particles (a computationally expensive, $N^2$
procedure which is not guaranteed to remove all fluctuations), as well
as spending time at the beginning of a simulation to evolve the system
in the presence of strong damping, while maintaining the desired
profiles of density, velocity, energy,
etc. \citep{2004MNRAS.351.1121R,2006MNRAS.365..199M,2009NewAR..53...78R}.

\begin{figure*}
    \begin{tabular}{c}
      \includegraphics[width=6in]{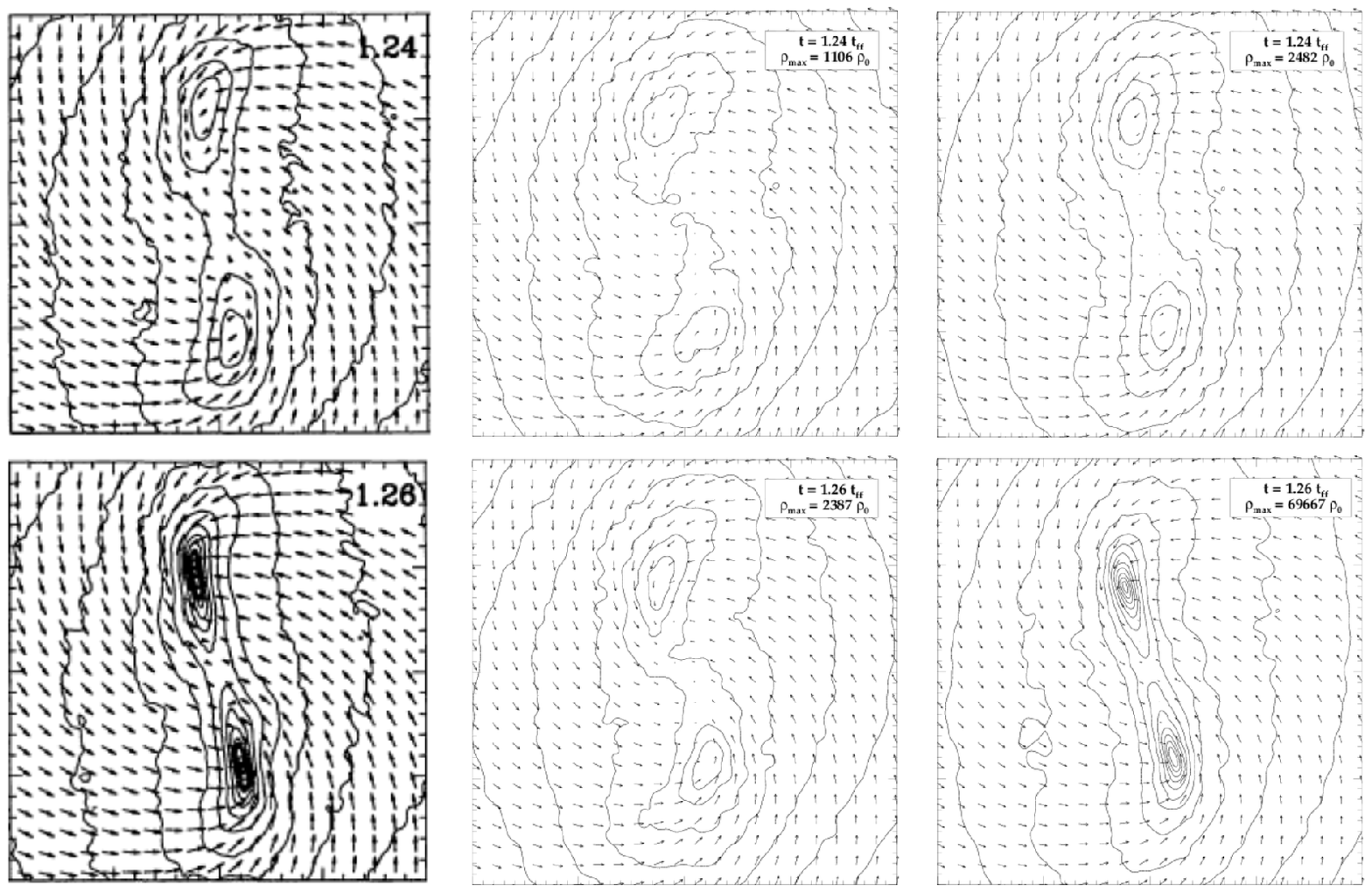}
   \end{tabular}
  \caption{\label{fig:sphere3sar} Isothermal sphere test (equatorial)
    density contours with four density contours are drawn per
    decade. Col. 1: `solution' from Fig.~4 of
    \citet{1997MNRAS.288.1060B}, for which $N=8\times10^4$; peak
    densities at $t=1.24$ and 1.26 are $\rho_{{\rm
        max}}/\rho_0\approx2,400$ and 78,500, respectively.  Cols. 2
    and 3: NEM and SAR distribution simulations, respectively; in
    both, $N=6.7\times10^4$.  SAR results show much greater agreement
    with the first column `solution', both qualitatively and
    quantitatively.}
\end{figure*}

In this work, we have utilised a new method to create equal-mass
particle distributions which are both smooth and random (SAR) from the
start of a simulation, i.e., limiting numerical fluctuations due to
initial placement. The SAR method produces such conditions at $t=0$,
without the need for special runtime considerations (such as damping,
etc.) during the simulation.  There is very little fixed cost in
creating such a distribution, and neither damping nor further
non-hydrodynamic numerics are introduced into the model.  The SAR
method for setting up initial conditions is valid for any number of
dimensions and does not assume any particular symmetries.

\subsection{SAR process steps}
\label{sec:sarprocess}
To make SAR initial conditions for a simulation, an intermediate
distribution of point particles is first created, called a `glass'.
The glass is defined by the property that its constituent particles
are equidistant from their nearest neighbours but without regular
structure (such as in a crystal).  For practical considerations of
generating a glass, it is computationally efficient to first make a
single, small glass distribution and then to tile copies with
periodic boundaries appropriately matched to preserve interparticle
spacing.

The general process of creating an SAR particle distribution may be divided
into the following steps:
\begin{enumerate}
 \item seed a mini-glass distribution: place a small number,
   $N_{{\rm g}}$, of point particles randomly in a cube of
   normalised edge length, $L_{{\rm g}}=1$, with periodic
   boundaries
 \item create the mini-glass: evolve the particles with mutually
   repulsive (here, inverse-square) forces, asymptotically approaching
   a constant density of equidistant particles (particle velocity is
   zeroed after each timestep to limit `overshooting' equilibrium)
 \item create full glass: tile copies of the mini-glass cubes (into
   any given size/shape), with the periodic boundaries of tiles
   matched to preserve particle spacing
 \item fill simulation model: select a small volume of the glass
   (including particles) and map it into the model, adjusting the
   element volume (and relative particle locations therein) to the
   local density of the profile by compressing or stretching it
 \item repeat the previous step, smoothly mapping neighbouring volumes
   from the glass to neighbouring volumes in the model profile,
   maintaining an absence of spurious particle number density
   fluctuations.
 \end{enumerate}
Using small volumes, such a  process preserves the relative spacing of
all  points while creating  an appropriate  density profile.

\subsection{Testing the SAR distributions}
\label{sartest}

Here, we first tested a SAR distribution model using a standard SPH
test of particle resolution and convergence: the collapse of a
self-gravitating, isothermal sphere initially in solid body rotation
and with a non-axisymmetric ($m=2$) density perturbation.  The system
fragments, and this has been well-examined using both Lagrangian and
various Eulerian grid codes
\citep{1979ApJ...234..289B,1997MNRAS.288.1060B,1998ApJ...495..821T}.
SAR models were tested with total particle number $N=3.4\times10^4$,
$6.7\times10^4$ and $2\times10^5$, and results were compared with
lattice-like, nonequal mass (NEM) placements using a standard
configuration (e.g. \citet{2005MNRAS.364.1105S}) at similar
resolutions.  Results of both simulations were also compared with
convergent `solution' results of \citet{1997MNRAS.288.1060B}.

Briefly, the initial conditions of the test are: at $t=0$, we start
with an isothermal ($P=c_{\rm{s}}^2\rho$) sphere having uniform sound
speed, $c_{\rm{s}}=1.66\times10^4~\rm{cm~s}^{-1}$; radius, $R_{\rm
  S}=5\times10^{16}~\rm{cm}$; mass, $M=1{\rm M}_{\odot}$; and solid
body angular velocity, $\omega=7.2\times10^{-13}~\rm{rad~s}^{-1}$.
The density distribution has an $m=2$ perturbation over the azimuthal
angle given by $\rho(\phi)=\rho_0[1+0.1\,\cos(2\phi)]$, where
$\rho_0=3.82\times10^{-18}~\rm{g~cm}^{-3}$.  All quantities in this
section are scaled by the following characteristic quantities: the
initial value of the outer radius, $R_{\rm S}$; the free-fall time,
$t_{{\rm ff}}=(3\pi/32G\rho_0)^{1/2}$; and the mass, ${\rm
  M}_{\odot}$. 

Fig.~\ref{fig:sphere3sar} shows plots of density in the equatorial
plane of the SAR and NEM distributions with $N=6.7\times10^4$
particles, as well as the convergent `solution' (top), at $t=1.24$ and
1.26. Quantitatively, the peak density of the SAR simulation was much
closer to the expected value than that of the NEM case (and at fairly
low resolution).  This remained true for all $N$ values investigated,
and furthermore, the SAR contours were significantly smoother and more
`fluidlike' with fewer numerical fluctuations.  The simulation results
using SAR conditions converged much more quickly to the expected
results as resolution was increased, an important property for
creating efficient simulations.

\begin{figure*}
    \begin{tabular}{l}
      \includegraphics[height=2.25in]{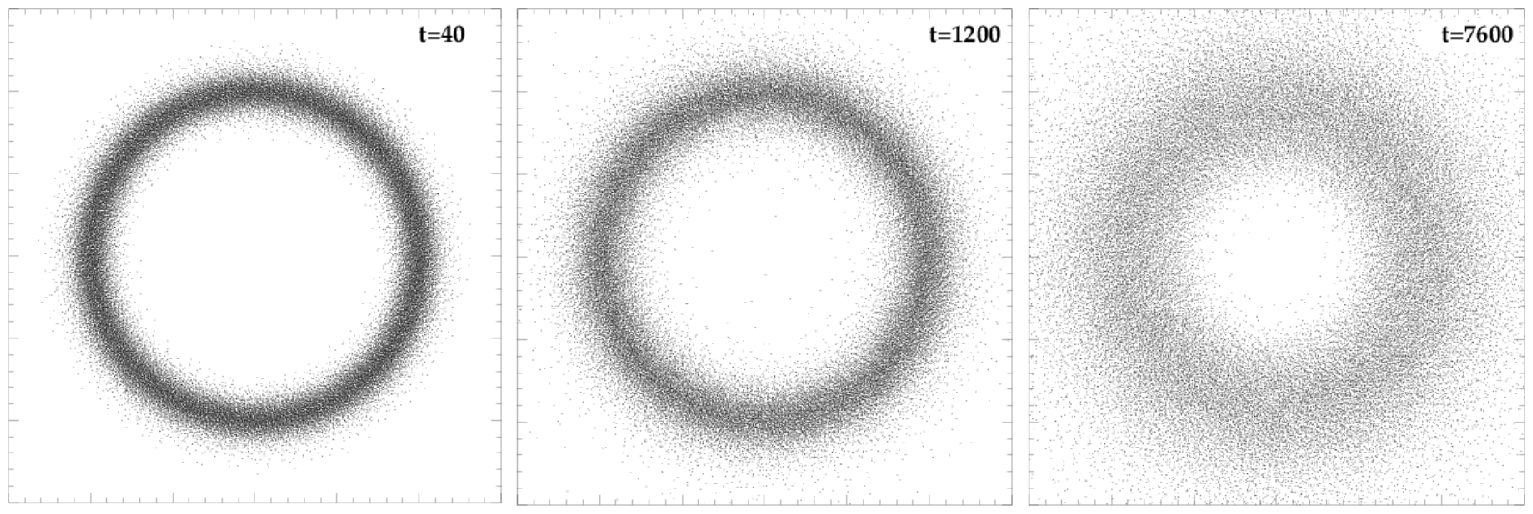}
  \end{tabular}
  \caption{\label{fig:ring_init} A sequence of positions of SPH
    particles, projected onto the equatorial plane, for the $\mu_1=60,
    N=5\times10^4$ model shown in the left column of
    Fig.~\ref{fig:ringsp_cs60}.  Note the early and continued absence
    of clumps, voids and temporary spiral.}
\end{figure*}

As an additional example of SAR results, Fig.~\ref{fig:ring_init}
shows images of particle locations for an evolving SAR ring simulation
(the same system as in the left panel of Fig.~\ref{fig:ringsp_cs60}),
projected onto the equatorial plane. Over time, the particle
distribution within system retains its good qualitative properties of
smoothness and randomness. Note the absence of any temporary
artificial spirals, clumps or sub-structures, such as often appear in
SPH simulations, even at very early times. Both the efficiency of
modelling and the relative ease of formulation (the computation
requires only a few simple loops through $N$ particles), in addition
to a reduction of numerical artefacts, make the SAR initial conditions
very beneficial for use in SPH simulations.

Importantly, the SAR particle method is effective because the
quasi-continuous mapping retains the relative spacing between
particles from the initial glass\footnote{A version of particle setup
  obtained by warping a constant density profile
  \citep{1994MmSAI..65.1013H} has been pointed out by a
  referee. Although the mapping feature therein is ostensibly similar
  to what is described here, that method was limited for use with a
  certain subset of spherically symmetric profiles and was made from
  uniformly placed (not SAR) particles.}.  The new density profile is
obtained smoothly and accurately, without the presence of fluctuations
or numerical artefacts. How smoothly a density profile is reproduced
is principally determined by the size of the incremental volume
element.  The SAR method does result in a small amount of localised
asymmetry in the particle distribution, as volume elements are not
generally scaled uniformly in all directions, but for small elements
this did not appear to produce spurious artefacts during simulations.
Improvements to the method are currently being considered.

\end{document}